\documentclass[a4paper,11pt]{article}
\pdfoutput=1 % if your are submitting a pdflatex

\usepackage{jheppub}

\usepackage[T1]{fontenc} % if needed

\usepackage{stmaryrd}

\newcommand{\su}{\mathfrak{su}}
\newcommand{\psu}{\mathfrak{psu}}

\newcommand{\gen}[1]{\mathbf{#1}}
\newcommand{\com}[2]{\big[#1,#2\big]}
\newcommand{\acomm}[2]{\big\{#1,#2\big\}}

\newcommand{\ket}[1]{\left|#1\right\rangle}

\newcommand{\bra}[1]{\left\langle#1\right|}

\newcommand{\MA}{\mathcal{A}}
\newcommand{\MB}{\mathcal{B}}
\newcommand{\MC}{\mathcal{C}}
\newcommand{\MD}{\mathcal{D}}

\newcommand\J{\star}

\renewcommand\L{\text{\tiny L}}
\newcommand\R{\text{\tiny R}}
\renewcommand\o{{\color{blue}\varobar}}
\newcommand\op{{\color{red}\varominus}}

\renewcommand\O{\circ}

\newcommand\B{\text{\tiny B}}
\newcommand\F{\text{\tiny F}}

\newcommand\LL{{\L\L}}
\newcommand\LR{{\L\R}}
\newcommand\Lo{{\L\o}}
\newcommand\Lop{{\L\op}}
\newcommand\RL{{\R\L}}
\newcommand\RR{{\R\R}}

\newcommand\oop{{\o\op}}

\newcommand\opo{{\op\o}}

\newcommand\LO{{\L\O}}
\newcommand\OL{{\O\L}}
\newcommand\RO{{\R\O}}
\newcommand\OR{{\O\R}}
\newcommand\OO{{\O\O}}

\usepackage[bb=boondox]{mathalfa}
\usepackage{mathbbol}
\usepackage{amssymb}
\DeclareSymbolFontAlphabet{\amsmathbb}{AMSb}

\newcommand{\Mono}{\mathbf{M}}
\newcommand{\Tran}{\mathbf{T}}
\newcommand{\MonoB}{\amsmathbb{M}}
\newcommand{\TranB}{\amsmathbb{T}}

\title{\boldmath Transfer matrices for AdS3/CFT2}

\author[a]{Fiona~K.~Seibold,}
\author[b,c,1]{Alessandro Sfondrini%
\note{IBM Einstein Fellow.}}

\affiliation[a]{Blackett Laboratory, Imperial College London,\\
Prince Consort Road, London, SW7 2AZ, UK}
\affiliation[b]{Institute for Advanced Study\\
Einstein Drive,
Princeton, New Jersey, 08540 USA }
\affiliation[c]{Dipartimento di Fisica e Astronomia, Universit\`a degli Studi di Padova,\\
\& Istituto Nazionale di Fisica Nucleare, Sezione di Padova,\\
via Marzolo 8, 35131 Padova, Italy}

\emailAdd{f.seibold21@imperial.ac.uk}
\emailAdd{alessandro.sfondrini@unipd.it}

\abstract{We work out the algebraic Bethe ansatz for the worldsheet theory of the $AdS_3\times S^3\times T^4$ superstring, and use it to derive the transfer matrices for fundamental particles and bound states of the string and  mirror model.
We also show how the Bethe equations and transfer matrices are modified in the presence of an Abelian twist.
These will be an important ingredient in the exploration of the mirror thermodynamic Bethe ansatz equations recently proposed by Frolov and Sfondrini, and their generalisation to twisted and deformed models.
}

\begin{document} 
\maketitle
\flushbottom
\newpage

%%%%%%%%%%%%%%%%%

\section{Introduction}
\label{sec:intro}

The study of the AdS3/CFT2 correspondence on the string worldsheet has recently seen a revival. In the cases where the geometry is realised without any Ramond-Ramond (RR) background fluxes, several exact computations can be efficiently performed by CFT techniques~\cite{Maldacena:2000hw}.
Things are more complicated in the presence of RR fluxes. There, it seems much harder to extract exact predictions from the theory, even when employing fairly sophisticated approaches~\cite{Berkovits:1999im}. On the other hand, the theory is classically integrable in the presence of arbitrary combinations of RR and Neveu-Schwarz-Neveu-Schwarz (NSNS) fluxes~\cite{Cagnazzo:2012se}. This raises hope that integrability may allow for a complete understanding \textit{at least} of the planar spectrum of the theory, similarly to what happened for $AdS_5\times S^5$ superstrings~\cite{Arutyunov:2009ga,Beisert:2010jr}. We refer the reader to~\cite{Sfondrini:2014via} for an overview of the early developments in the integrability approach to AdS3/CFT2.

Very recently, the thermodynamic Bethe ansatz (TBA) equations for ``mirror model''~\cite{Arutyunov:2007tc} of the $AdS_3\times S^3\times T^4$ superstring (which is the prototypical AdS3/CFT2 setup) have been formulated in~\cite{Frolov:2021bwp}.%
\footnote{%
More specifically, the mirror TBA of~\cite{Frolov:2021bwp} was formulated for pure-RR setup. The case of pure-NSNS backgrounds, which was already understood by worldsheet CFT, can also be described by the mirror TBA~\cite{Baggio:2018gct,Dei:2018mfl,Dei:2018jyj}. The mixed-flux case would require new insights on the dressing factors of the model, most likely building on the recent proposal of~\cite{Frolov:2021fmj}.
} 
These equations are meant to exactly capture the finite-volume spectrum of the theory, unlike the Bethe-Yang equations (which instead only capture the ``asymptotic'' spectrum, up to the so-called wrapping corrections). Still, to study in detail the mirror TBA it is important to have a detailed understanding of the asymptotic model, in the string and mirror theory. One important ingredient are the features of the so-called \textit{Y-functions}, which are central object to be determined by solving the mirror TBA equations. Typically, these features are determined starting from the large-volume theory, where the asymptotic solution can be qualitatively trusted, see \textit{e.g.}~\cite{Bajnok:2008bm,Gromov:2009tv,Arutyunov:2009ax} for the case of $AdS_5\times S^5$. In practice, understanding the asymptotic Y-functions requires knowing the eigenvalues of the transfer matrix for fundamental particles and bound states.

Our main motivation is therefore to work out these eigenvalues, both in the direct and mirror model. While the Bethe-Yang equations for the $AdS_3\times S^3\times T^4$ model are well-known, having been derived in~\cite{Borsato:2014hja} based on~\cite{Borsato:2012ss}, the transfer matrix has not yet been studied in complete detail.
One reason is that the Bethe-Yang equations were originally derived in the ``nested'' approach, \textit{cf.}~\cite{Beisert:2005tm}, which is less convenient for deriving the eigenvalues of the transfer matrix. For this purpose, the algebraic Bethe ansatz is a better choice, and it is what we shall use here.

The algebraic Bethe ansatz for $AdS_3\times S^3\times T^4$ was also discussed, to a limited extent, in~\cite{Majumder:2021zkr}. There, the authors were mostly focusing on the protected states of this model~\cite{Borsato:2016kbm,Baggio:2017kza}, and therefore carried out their construction only in one sector of the theory (the one where all particles are massless). Here, we are interested in a more general discussion of the model, both in the string and mirror theory. Note also that the transfer matrices of AdS3/CFT2 were studied recently in~\cite{DeLeeuw:2020ahx} through a map to free Fermions. The presentation of those results was rather implicit, and we feel that it is appropriate to work out the full structure of the transfer matrix of this~model.

The derivation of the algebraic Bethe ansatz for models of this type is by now well-understood~\cite{Faddeev:1996iy,Martins:2007hb}. There are however a few peculiar points for the $AdS_3\times S^3\times T^4$ superstring that are worth investigating in some detail. First of all, the fundamental particles of the theory transform in \textit{several} irreducible representations of the symmetry algebra~\cite{Borsato:2013qpa}. As we shall see below, these are called the ``left'', ``right'' and ``massless'' representations. In the algebraic Bethe ansatz, we want to construct eigenstates of the transfer matrix by acting with a suitable creation operator, conventionally called~$\mathcal{B}(y)$, where $y$ is some suitable rapidity which parametrises the operator's action. Considering the several irreducible representations, we have the possibility to act with several types of $\mathcal{B}$ operator, like $\mathcal{B}_\L(y_\L)$ in the left and $\mathcal{B}_\R(y_\R)$ in the right. In order to construct the states it is important to check whether this action always results in \textit{distinct} states or not. This is crucial to correctly account for the degeneracy of states, which in turn is an important ingredient in deriving the TBA equations. We will show that, in fact, the action of these two creation operators is not independent. In other words, if $y_\R(y_\L)$  is a suitable function of $y_\L$, then the states created by~$\mathcal{B}_\L(y_\L)$ and~$\mathcal{B}_\R(y_\R(y_\L))$ are proportional, and should only be counted once. This was already argued in~\cite{Borsato:2012ss} based on the nested Bethe ansatz construction, but the more rigorous framework of the algebraic Bethe ansatz allows us to present a more convincing~proof.

Another point which has not been discussed much in the literature is the derivation of the Bethe equations and transfer matrices for the string \textit{viz.}\ mirror bound states of the theory.%
\footnote{The Bethe equations for mirror bound states were recently derived in~\cite{Frolov:2021bwp} as an intermediate step in the derivation of the mirror TBA.}
 As we will review, this model can have ``left'' bound states and ``right'' bound states. With respect to $AdS_5\times S^5$ the bound-state scattering is substantially simpler because the dimension of the bound-state representation does not grow with the bound-state number~\cite{Borsato:2013hoa}. As a result, it is not necessary to work out a generating function of the transfer matrix like \textit{e.g.}\ in $AdS_5\times S^5$~\cite{Arutyunov:2009iq} (though it might be interesting to revisit that construction for $AdS_3 \times S^3 \times T^4$), and the derivation of the bound-state transfer matrix is fairly explicit both in the  string and mirror models. These two models are related by analytic continuation~\cite{Arutyunov:2007tc, Frolov:2021zyc}. Moreover, the  string and mirror bound-states sit in inequivalent representations (this is also the case in $AdS_5\times S^5$~\cite{Arutyunov:2007tc}). We will see that we can account for this last fact by swapping the grading of the left- and right-representation when going to the mirror model.

Having understood the transfer matrix of the string and mirror model, it is relatively straightforward to extend this discussion to the \textit{twisted} transfer matrix. In this case, instead of obtaining the transfer matrix as the (super)trace of the monodromy matrix, we consider a twisted trace, where the twist is a generic element of the Abelian subalgebra of the symmetry algebra of the theory. Physically, this twisted transfer matrix captures the physics of orbifolded and TsT-transformed backgrounds~\cite{Lunin:2005jy}, see \textit{e.g.}~\cite{vanTongeren:2013gva} for a review.

This paper is structured as follows. In section~\ref{sec:particles} we briefly review the particle content and symmetries of the model. In section~\ref{sec:BAonecopy} we discuss the building blocks for the algebraic Bethe ansatz construction (exploiting the factorised structure of the symmetry algebra), which we use in section~\ref{sec:BAfull} to derive the full algebraic Bethe ansatz and transfer matrices of the full model. In section~\ref{sec:twists} we consider the most general model twisted with Abelian twists. We discuss the outlook of this work in section~\ref{sec:conclusions}. Finally, in appendix~\ref{app:boundstates} we collect some useful observations about bound states and in appendix~\ref{app:Null} we show how to correctly identify the independent auxiliary excitations of the model.

%%%%%%%%%%%%%%%%%

\section{Particles and Representations}
\label{sec:particles}

We begin by reviewing the particle content of the worldsheet theory for $AdS_3\times S^3\times T^4$. As discussed in~\cite{Borsato:2013qpa,Borsato:2014exa}, the fundamental particles of the model fall into four irreducible representations of $\psu(1|1)^{\oplus4}$ centrally extended. The representations are four-dimensional, for a total of 16 fundamental particles, as expected for strings in lightcone gauge. As we shall see, it is actually convenient to build these irreducible representations out of tensor products of \textit{two-dimensional} representations of a smaller algebra. This is analogous to how representations of the fundamental excitations of $AdS_5\times S^5$ are constructed, with the important difference that here we deal with \textit{several} irreducible representations, rather than~one.

\subsection{Lightcone symmetry algebra and central charges}
\label{sec:particles:algebra}
The algebra is given by
\begin{equation}
\label{eq:bigalgebra}
\begin{aligned}
\acomm{\gen{Q}^\alpha}{\gen{S}_\beta} &= \delta^\alpha_\beta\,\gen{H}\,,\qquad&
\acomm{\widetilde{\gen{Q}}_\alpha}{\widetilde{\gen{S}}^\beta} &= \delta_\alpha^\beta\,\widetilde{\gen{H}}\,,\\
\acomm{\gen{Q}^\alpha}{\widetilde{\gen{Q}}_\beta} &= \delta^\alpha_\beta\,\gen{C}\,,\qquad&
\acomm{\gen{S}_\alpha}{\widetilde{\gen{S}}^\beta} &= \delta_\alpha^\beta\,\overline{\gen{C}}\,,
\end{aligned}
\end{equation}
with $\alpha=1,2$ and $\beta=1,2$. We are interested in the unitary short irreducible representation of this algebra, which are four dimensional and obey the conditions
\begin{equation}
\label{eq:shortening}
    \gen{H}\,\widetilde{\gen{H}}=\gen{C}\,\overline{\gen{C}}\,,\qquad
    \gen{S}_\alpha=\big(\gen{Q}^\alpha\big)^\dagger\,,
    \quad\widetilde{\gen{S}}^\alpha=\big(\widetilde{\gen{Q}}_\alpha\big)^\dagger\,.
\end{equation}
Following this, it is convenient to introduce the lightcone energy~$\gen{E}$ and the $\mathfrak{u}(1)$ charge $\gen{M}$
\begin{equation}
    \gen{E}=\gen{H}+\widetilde{\gen{H}}\geq0\,,\qquad
    \gen{M}=\gen{H}-\widetilde{\gen{H}}\,.
\end{equation}

It turns out that the eigenvalue of $\gen{C}$ (and thus of $\overline{\gen{C}}=\gen{C}^*$) depends only on the parameters of the theory, and not on the type of particle. In particular we have~\cite{Borsato:2014hja}
\begin{equation}
    \overline{\gen{C}}^*=\gen{C}=\frac{ih}{2}(e^{ip}-1)\,\gen{1}\,.
\end{equation}
Here $h\geq0$ is the amount of RR flux of the theory, and $p$ is the worldsheet momentum of the particle.
It is the eigenvalue of $\gen{M}$ that distinguishes different types of particles, namely
\begin{equation}
    \gen{M}= m\,\gen{1}\,,\qquad m\in\amsmathbb{Z}\,.
\end{equation}
Here~$m$ labels different types of particles (it is the sum of the spin on AdS and on the three-sphere~\cite{Sfondrini:2014via}). In particular we have%
\footnote{%
The eigenvalues of $m$ would take different values for theories supported by a mixture of RR and NSNS fluxes~\cite{Hoare:2013lja,Lloyd:2014bsa}.
}
\begin{itemize}
    \item One ``left'' representation with $m=+1$, containing four particles which we call $Z,Y$ (a Boson in AdS and one on the sphere, respectively) and $\Psi^{\alpha}$.%
\footnote{The index~$\alpha$ is the same labeling the supercharges, and it corresponds to an $\su(2)$ automorphism, so-called $\su(2)_\bullet$.}
    \item One ``right'' representation with $m=-1$, containing four particles which we call $\tilde{Z},\tilde{Y}$ (a Boson in AdS and one on the sphere, respectively) and $\tilde{\Psi}^{\alpha}$.
        \item Two ``massless'' representations with $m=0$, distinguished by an additional index~$\dot{\alpha}$. Altogether, they contain four $T^4$ Bosons~$T^{\dot{\alpha}\alpha}$ and four Fermions~$\chi^{\dot{\alpha}}$, $\tilde{\chi}^{\dot{\alpha}}$.
\end{itemize}
Additionally, it is possible to construct bound-state representations. They come in two families:
\begin{itemize}
    \item ``Left'' bound-state representations, with $m=+2,+3,\dots$, containing two Bosons and two Fermions each.
    \item ``Right'' bound-state representations, with $m=-2,-3,\dots$, containing two Bosons and two Fermions each.
\end{itemize}
These representations can also be constructed from taking the tensor product of fundamental-particle representations and imposing suitable constraints in a way which we will detail below. The construction can be performed both for the string theory and the mirror theory, though it is not identical.

\subsection{Factorised structure and representations}
\label{sec:particles:representations}
The representations which we want to consider can be obtained from considering a smaller, algebra, namely~\cite{Borsato:2012ud}
\begin{equation}
\label{eq:smallalgebra}
\begin{aligned}
    \acomm{\gen{q}}{\gen{s}}&=\gen{H}\,,\qquad&
    \acomm{\tilde{\gen{q}}}{\tilde{\gen{s}}}&=\widetilde{\gen{H}}\,,\\
    \acomm{\gen{q}}{\tilde{\gen{q}}}&=\gen{C}\,,\qquad&
    \acomm{\gen{s}}{\tilde{\gen{s}}}&=\overline{\gen{C}}\,,
\end{aligned}
\end{equation}
which is related to~\eqref{eq:bigalgebra} by setting
\begin{equation}
\begin{aligned}
    \gen{Q}^1=\gen{q}\otimes\gen{1}\,,\qquad
    \gen{Q}^2=\Sigma\otimes\gen{q}\,,\qquad
    \gen{S}_1=\gen{s}\otimes\gen{1}\,,\qquad
    \gen{S}_2=\Sigma\otimes\gen{s}\,,\\
    \widetilde{\gen{Q}}_1=\tilde{\gen{q}}\otimes\gen{1}\,,\qquad
    \widetilde{\gen{Q}}_2=\Sigma\otimes\tilde{\gen{q}}\,,\qquad
    \widetilde{\gen{S}}^1=\tilde{\gen{s}}\otimes\gen{1}\,,\qquad
    \widetilde{\gen{S}}^2=\Sigma\otimes\tilde{\gen{s}}\,,
\end{aligned}
\end{equation}
where $\Sigma=(-\gen{1})^{\gen{F}}$ is the Fermion sign.

The short representations of the algebra~\eqref{eq:smallalgebra}, which also satisfy the shortening condition~\eqref{eq:shortening}, are two-dimensional.
We need the following types of representations, which we denote as (highest-weight$|$lowest-weight):
\begin{equation}
\rho_\L = ( \phi_\L | \varphi_\L)\,, \qquad \rho_\R = (\phi_\R | \varphi_\R)\,,
\qquad
\rho_\circ = ( \phi_\circ | \varphi_\circ)\,.
\end{equation}
Here the first state~$\phi$ is the highest-weight state, and the second~$\varphi$ is the lowest-weight one. In some cases, we will take~$\phi$ to be a Boson (and $\varphi$ a Fermion), or viceversa. When this distinction will be important we will add labels B and F as appropriate. Notice that changing the statistics of the highest-weight state results in an inequivalent superalgebra representation.
The representations of type $\rho_\L$ have $m=+1,+2,+3,\dots$, those of type $\rho_\R$ have $m=-1,-2,-3,\dots$, and those of type $\rho_\circ$ have~$m=0$.

\paragraph{Left representation.}
The explicit form of the representation in the basis $( \phi_\L, \varphi_\L)$ is%
\footnote{%
The subscripts ``L'' and, below, ``R'' indicate which (matrix) representation we are considering.}
\begin{equation}
\label{eq:representationL}
    \gen{q}_\L=\left(
    \begin{array}{cc}
         0&0  \\
         a&0 
    \end{array}
    \right)\,,\qquad
    \gen{s}_\L=\left(
    \begin{array}{cc}
         0&\bar{a}  \\
         0&0 
    \end{array}
    \right)\,,\qquad
    \tilde{\gen{s}}_\L=\left(
    \begin{array}{cc}
         0&0  \\
         \bar{b}&0 
    \end{array}
    \right)\,,\qquad
    \tilde{\gen{q}}_\L=\left(
    \begin{array}{cc}
         0&b  \\
         0&0 
    \end{array}
    \right)\,,
\end{equation}
where the coefficients depend on the momentum~$p$ and satisfy $(a(p))^*=\bar{a}(p^*)$, and similarly for $b$. We will define the representation parameters below. 
For fundamental particles in string theory, we will be interested in the case in which $\phi_\L$ is a~Boson.

\paragraph{Right representation.}
In the basis $( \phi_\R, \varphi_\R)$ the representation is
\begin{equation}
\label{eq:representationR}
    \gen{q}_\R=\left(
    \begin{array}{cc}
         0&0  \\
         b&0 
    \end{array}
    \right)\,,\qquad
    \gen{s}_\R=\left(
    \begin{array}{cc}
         0&\bar{b}  \\
         0&0 
    \end{array}
    \right)\,,\qquad
    \tilde{\gen{s}}_\R=\left(
    \begin{array}{cc}
         0&0  \\
         \bar{a}&0 
    \end{array}
    \right)\,,\qquad
    \tilde{\gen{q}}_\R=\left(
    \begin{array}{cc}
         0&a  \\
         0&0 
    \end{array}
    \right)\,,
\end{equation}
which differs from the left representation by the swap $a\leftrightarrow b$, which amounts to swapping the role of left- and right-supercharges. For fundamental particles in the string theory we will be interested in the case where $\phi_\R$ is a~Fermion.

\paragraph{Massless representation.}
The massless representation, in the basis $( \phi_\circ, \varphi_\circ)$ can be obtained either from~\eqref{eq:representationL} or~\eqref{eq:representationR}. In fact, by using the explicit form of $a,b,\bar{a}$ and~$\bar{b}$ which we give just below, it is possible to show that when~$m=0$ the limit of~\eqref{eq:representationL} is isomorphic to the limit of~\eqref{eq:representationR}. The isomorphism is simply the rescaling of say~$\varphi$ by a sign~\cite{Borsato:2014hja}.
To describe the massless particles of the string theory we need both the case in which $\varphi_\circ$ is a Boson and that when it is a~Fermion.

\paragraph{Representation coefficients.}
The representation coefficients, which implicitly depend on the mass~$m$, are
\begin{equation}
\begin{aligned}
    a&=\eta\,e^{i\xi}\,,\qquad&
    \bar{a}&=e^{-\frac{i}{2}p}\eta\,e^{-i\xi}\,,\\
    b&=-\frac{e^{-\frac{i}{2}p}}{x^-}\eta\,e^{i\xi}\,,\qquad&
    \bar{b}&=-\frac{1}{x^+}\eta\,e^{-i\xi}\,,
\end{aligned}
\end{equation}
where~$\xi$ is a representation parameter useful to build the two-particle representations~\cite{Arutyunov:2009ga}, and we introduced the functions $\eta=\eta(p,m,h)$ and $x^\pm(p,m,h)$ (we will often omit the dependence on the arguments). We have
\begin{equation}
    \eta=e^{\frac{i}{4}p}\sqrt{\frac{ih}{2}\big(x^- - x^+\big)}\,,
\end{equation}
with
\begin{equation}
    x^\pm= \frac{|m|+\sqrt{m^2+4h^2\sin^2(p/2)}}{2h\,\sin(p/2)}\,e^{\pm \frac{i}{2}p}\,.
\end{equation}
In particular, using these expressions, we find that the energy takes the form
\begin{equation}
    E=a\bar{a}+b\bar{b}=\frac{h}{2i}\left(x^+ - \frac{1}{x^+}-x^- + \frac{1}{x^-}\right)=\sqrt{m^2+4h^2\sin^2(p/2)}\,,
\end{equation}
and the mass
\begin{equation}
    m=\pm(a\bar{a}-b\bar{b}) = \pm\frac{h}{2i}\left(x^+ + \frac{1}{x^+}-x^- - \frac{1}{x^-}\right)\,,
\end{equation}
where in the two last equalities we pick the plus sign for left representations and the minus sign for right representations. Note that for $m=0$ we have $a\bar{a}=b\bar{b}$ as it must be.
We also have
\begin{equation}
    C=
    e^{2i\xi}\frac{ih}{2}\left(\frac{x^+}{x^-}-1\right)=
    e^{2i\xi}\frac{ih}{2}\left(e^{ip}-1\right)\,,
\end{equation}
and $\overline{C}=C^*$. We see that in the one-particle representation we may set~$\xi=0$. 

\paragraph{Fundamental particles.}
It is worth to briefly review how the fundamental particles, which tranform under $\psu(1|1)^{\oplus4}$, see~\eqref{eq:bigalgebra}, arise from the various~$\psu(1|1)^{\oplus2}$ representations constructed above. We have that the four ``left'' particles with $m=+1$ come from
\begin{equation}
\label{eq:psu114module}
    m=+1:\qquad(\phi^\B_\L|\varphi^\F_\L)\otimes(\phi^\B_\L|\varphi^\F_\L)\,.
\end{equation}
Here B and F denote Bosons and Fermions, respectively. The highest-weight state, $\phi^\B_\L\otimes \phi^\B_\L$, is a Boson on the three-sphere; the lowest weight state, $\varphi^\F_\L\otimes\varphi^\F_\L$ is a Boson on AdS; the two remaining states are Fermions. It is easy to verify that this is indeed a four-dimensional short irreducible representation of $\psu(1|1)^{\oplus4}$, see~\eqref{eq:bigalgebra}.
Similarly, for right particles we have
\begin{equation}
    m=-1:\qquad(\phi^\F_\R|\varphi^\B_\R)\otimes(\phi^\F_\R|\varphi^\B_\R)\,.
\end{equation}
Here the highest-weight state is an AdS Boson, namely $\phi^\F_\R\otimes\phi^\F_\R$, and the lowest-weight state is a sphere Boson. Finally, we have \textit{eight} massless particles, transforming in two irreducible representations:
\begin{equation}
    m=0:\qquad
    \Big[(\phi^\B_\circ|\varphi^\F_\circ)\otimes(\phi^\F_\circ|\varphi^\B_\circ)\Big]
    \oplus
    \Big[(\phi^\F_\circ|\varphi^\B_\circ)\otimes(\phi^\B_\circ|\varphi^\F_\circ)\Big]\,.
\end{equation}
The highest- and lowest-weight states here are Fermions, and the four Bosons are the coordinates of the four-torus. We will see that similar patterns are valid for bound states, with $|m|=2,3,4,\dots$.

\subsection{Bound states in the string and mirror theories}
%%%%%%%%%%%%%%%%%
In order to discuss the bound states of the theory it is first necessary to describe multi-particle representations. Here some care is needed because we expect the multi-particle eigenvalue of~$C$ and~$\overline{C}$ to be~\cite{Borsato:2014exa}
\begin{equation}
    C=\frac{ih}{2}\left(e^{i(p_1+\cdots +p_n)}-1\right)\,,
\end{equation}
and $\overline{C}=C^*$. This can be achieved by appropriately choosing the representation parameters $(\xi_1,\dots \xi_n)$ or equivalently by introducing a non-trivial coproduct~\cite{Borsato:2014hja}.
For the two-particle representation we can set, making the dependence on~$(\xi_1,\xi_2)$ explicit,
\begin{equation}
\label{eq:twoparticlerepr}
        \gen{q}(p_1,p_2)=\gen{q}(p_1,\xi_1=1)\otimes\gen{1}+\Sigma\otimes\gen{q}(p_2,\xi_2=\tfrac{1}{2}p_1)\,,
\end{equation}
where~$\gen{q}$ is any of the supercharges in any of the $\psu(1|1)^{\oplus2}$ representations introduced above. For generic values of $p_1,p_2$, these are four-dimensional long irreducible representations $\psu(1|1)^{\oplus2}$.%
\footnote{These representations have nothing to do with the short representations of $\psu(1|1)^{\oplus4}$ introduced \textit{e.g.}~in~\eqref{eq:psu114module}, and should not be confused with them.}
However, for special (complex) values of~$p_1,p_2$, the representation may become reducible (but indecomposable), and reveal the existence of a short \textit{bound-state representation} of dimension two. 
Much like in $AdS_5\times S^5$~\cite{Arutyunov:2008zt}, the bound-state representation can be constructed by modding out the indecomposable representation (which here is four-dimensional) by a (two-dimensional) subrepresentation.The name ``bound state'' is justified not only by the complex momenta, but by the fact that there are poles in the S~matrix corresponding to such values of the momentum. Additionally, the existence of these bound states can be also verified semiclassically.

\paragraph{General structure of the bound-state representation.}
Let us start by writing an explicit base of the two-particle representation, starting from the long four-dimensional representation. We have
\begin{equation}
    \ket{\phi\otimes\phi}\,,\qquad
    \ket{\phi\otimes\varphi}\,,\qquad
    \ket{\varphi\otimes\phi}\,,\qquad
    \ket{\varphi\otimes\varphi}\,,
\end{equation}
where the first element of the list is the highest-weight state and the last is the lowest-weight state. There are two possibilities to find a short (two-dimensional) irreducible representation within this module. Firstly, observe that any two-dimensional irreducible representation must contain exactly one Boson, which may be either $\ket{\phi\otimes\phi}$ or $\ket{\varphi\otimes\varphi}$. Hence, the two choices are:
\begin{enumerate}
    \item The short representation contains $\ket{\phi\otimes\phi}$, which must be its highest-weight state, in which case the other linearly independent state is the descendant $\gen{q}\ket{\phi\otimes\phi}$.
    \item The short representation contains $\ket{\varphi\otimes\varphi}$, which must be its lowest-weight state, in which case the other linearly independent state is the ascendant $\gen{s}\ket{\varphi\otimes\varphi}$.
\end{enumerate}
We will see that we will need to pick either option depending on whether we are dealing with left or right bound states, and whether we are in the string or mirror theory.

\paragraph{Bound states in the string theory.}
The bound states in the string theory arise upon imposing the condition
\begin{equation}
    x^+(p_1)=x^-(p_2)\,.
\end{equation}
Let us firstly verify the value of the central charges. By using~\eqref{eq:twoparticlerepr} it is easy to see that
\begin{equation}
    \gen{E}_{12}=\gen{E}\otimes\gen{1}+\gen{1}\otimes\gen{E}\,,\qquad
    \gen{M}_{12}=\gen{M}\otimes\gen{1}+\gen{1}\otimes\gen{M}\,,
\end{equation}
meaning that energy and mass  are simply additive, and the parameter $\xi$ plays no role. Hence we find the energy
\begin{equation}
    E=E(p_1)+E(p_2)=\frac{h}{2i}\left(x^+(p_2) - \frac{1}{x^+(p_2)}-x^-(p_1) + \frac{1}{x^-(p_1)}\right)\,,
\end{equation}
and the mass
\begin{equation}
    m=m_1+m_2=\pm\frac{h}{2i}\left(x^+(p_2) + \frac{1}{x^+(p_2)}-x^-(p_1) - \frac{1}{x^-(p_1)}\right)\,,
\end{equation}
where again the sign depends on whether we are considering two left particles, or two right particles. There are no physical bound states involving one left and one right particle. For the remaining central charge
\begin{equation}
    C=
    \frac{ih}{2}\left(\frac{x^+(p_2)}{x^-(p_1)}-1\right)\,,
\end{equation}
and $\overline{C}=C^*$.
Clearly this representation has the same central charges as one of the representations constructed above with $p=p_1+p_2$, $m=m_1+m_2$, and
\begin{equation}
    x^+(p_2;m_1) = x^+(p,m)\,,\qquad
    x^-(p_1;m_1) = x^-(p,m)\,.
\end{equation}
This can also be nicely expressed on the $u$-rapidity plane, see~\cite{Frolov:2021fmj}. For our purposes, this is \textit{almost} sufficient to completely determine the representation that we need to consider. We just need to figure out whether the highest-weight state is Bosonic or Fermionic.
This can be directly verified by constructing the representation space, as we do in appendix~\ref{app:boundstates}. However, an easier way to reach the same conclusions is to recall that, physically, the bound states are related to the dynamics on the three-sphere~\cite{Borsato:2013hoa} much like in $AdS_5\times S^5$~\cite{Dorey:2006dq}. In other words:
\begin{enumerate}
    \item A bound state of two left particles should contain 
\begin{equation}
\ket{\phi^\B_\L\otimes\phi^\B_\L} 
\sim
\ket{\phi^\B_\L(p_1,p_2)}\,,    
\end{equation}    
and the other state must be the descendant
\begin{equation}
\gen{q}\ket{\phi^\B_\L\otimes\phi^\B_\L}
=a(p_1)\,\ket{\varphi^\F_\L\otimes\phi^\B_\L}
+a(p_2)\,\ket{\phi^\B_\L\otimes\varphi^\F_\L}\,.
\end{equation}
Hence this module is of the type $(\phi^\B_\L|\varphi^\F_\L)$.
\item A bound state of two right particles should contain 
\begin{equation}
\ket{\varphi^\B_\R\otimes\varphi^\B_\R}\sim 
\ket{\varphi^\B_\R(p_1,p_2)}\,,    
\end{equation}    
and the other state must be the ascendant
\begin{equation}
\gen{s}\ket{\varphi^\B_\R\otimes\varphi^\B_\R}
=\bar{a}(p_1)\,\ket{\phi^\F_\R\otimes\varphi^\B_\R}
+\bar{a}(p_2)\,\ket{\varphi^\B_\R\otimes\phi^\F_\R}\,.
\end{equation}
Hence this module is of the type $(\phi^\F_\R|\varphi^\B_\R)$.
\end{enumerate}
Because of the signs that appear in the linear combinations above, both of these choices are called \textit{symmetric} bound state representations.
In short, we have found that the  (left and right)  bound-state representations of the string theory have exactly the same form as the (left and right) fundamental-particle ones, up to changing the value of $m$ in the parametrisation of~$x^\pm$.

\paragraph{Bound states in the mirror theory.}
The kinematics of the mirror theory is rather different from that of the original theory, see~\cite{Arutyunov:2007tc, Frolov:2021zyc}. Still, it is possible to obtain the mirror representations and S~matrix from the analytic continuation of the original theory. Therefore, we can formally obtain much of the construction of the mirror theory from the one of the direct theory up to an analytic continuation.
One important difference between the mirror theory and the string theory is which bound states are allowed~\cite{Frolov:2021bwp}. In fact, in the mirror theory we have
\begin{equation}
    x^-(p_1)=x^+(p_2)\,.
\end{equation}
It is easy to see, as we discuss in appendix~\ref{app:boundstates}, that this condition selects a different subspace of the two-particle module with respect to the string theory. In particular, we find that
\begin{enumerate}
    \item For left particles, the bound state representation is of the form $(\phi^\F_\L|\varphi^\B_\L)$, with opposite grading with respect to the string theory.
    \item For right particles, the bound state representation is of the form $(\phi^\B_\R|\varphi^\F_\R)$, again with opposite grading with respect to the string theory.
\end{enumerate}

\subsection{S matrix in the string and mirror model}
\label{sec:particles:signs}
We have seen that, when going from the string to the mirror model, we should reverse the grading of the highest-weight state in the left and right representations. To avoid repeating the analysis of the transfer matrix in the two cases, it is worth determining how the S~matrix changes under this flip of statistics. Unsurprisingly, this boils down to flipping the sign of certain S-matrix elements.

\paragraph{S matrix from the symmetries.}
To begin with, let us recap the structure of the S~matrix, which is split off in left-left, right-right, left-right, and right-left blocks.%
\footnote{%
We can similarly also define the blocks involving massless particles, which can be obtained as limits of left particles. In that limit, the representations and the S~matrix depends only on $x^+$ while $x^-=1/x^+$.
} 
We will work with the two-dimensional representations of $\su(1|1)^{\oplus2}$ centrally extended, so that each block will act on a four-dimensional vector space. Each of these blocks, denoted by~$\gen{S}_{\L\L}(x_1^\pm,x_2^\pm), \gen{S}_{\R\R}(x_1^\pm,x_2^\pm), \gen{S}_{\L\R}(x_1^\pm,x_2^\pm)$, and $\gen{S}_{\R\L}(x_1^\pm,x_2^\pm)$ will have to obey the invariance conditions, which \textit{e.g.}\ for  LL read
\begin{equation}
\label{eq:llinvariance}
\begin{aligned}
    \gen{S}_{\L\L}(x_1^\pm,x_2^\pm)\,\gen{q}_{\L\L}(x_1^\pm,x_2^\pm)&=
    \gen{q}_{\L\L}(x_2^\pm,x_1^\pm)\,\gen{S}_{\L\L}(x_1^\pm,x_2^\pm)\,,\\
    \gen{S}_{\L\L}(x_1^\pm,x_2^\pm)\,\gen{s}_{\L\L}(x_1^\pm,x_2^\pm)&=
    \gen{s}_{\L\L}(x_2^\pm,x_1^\pm)\,\gen{S}_{\L\L}(x_1^\pm,x_2^\pm)\,,\\
    \gen{S}_{\L\L}(x_1^\pm,x_2^\pm)\,\tilde{\gen{q}}_{\L\L}(x_1^\pm,x_2^\pm)&=
    \tilde{\gen{q}}_{\L\L}(x_2^\pm,x_1^\pm)\,\gen{S}_{\L\L}(x_1^\pm,x_2^\pm)\,,\\
    \gen{S}_{\L\L}(x_1^\pm,x_2^\pm)\,\tilde{\gen{s}}_{\L\L}(x_1^\pm,x_2^\pm)&=
    \tilde{\gen{s}}_{\L\L}(x_2^\pm,x_1^\pm)\,\gen{S}_{\L\L}(x_1^\pm,x_2^\pm)\,,
\end{aligned}
\end{equation}
while for \textit{e.g.}\ LR they read
\begin{equation}
\label{eq:lrinvariance}
\begin{aligned}
    \gen{S}_{\L\R}(x_1^\pm,x_2^\pm)\,\gen{q}_{\L\R}(x_1^\pm,x_2^\pm)&=
    \gen{q}_{\R\L}(x_2^\pm,x_1^\pm)\,\gen{S}_{\L\R}(x_1^\pm,x_2^\pm)\,,\\
    \gen{S}_{\L\R}(x_1^\pm,x_2^\pm)\,\gen{s}_{\L\R}(x_1^\pm,x_2^\pm)&=
    \gen{s}_{\R\L}(x_2^\pm,x_1^\pm)\,\gen{S}_{\L\R}(x_1^\pm,x_2^\pm)\,,\\
    \gen{S}_{\L\R}(x_1^\pm,x_2^\pm)\,\tilde{\gen{q}}_{\L\R}(x_1^\pm,x_2^\pm)&=
    \tilde{\gen{q}}_{\R\L}(x_2^\pm,x_1^\pm)\,\gen{S}_{\L\R}(x_1^\pm,x_2^\pm)\,,\\
    \gen{S}_{\L\R}(x_1^\pm,x_2^\pm)\,\tilde{\gen{s}}_{\L\R}(x_1^\pm,x_2^\pm)&=
    \tilde{\gen{s}}_{\R\L}(x_2^\pm,x_1^\pm)\,\gen{S}_{\L\R}(x_1^\pm,x_2^\pm)\,.
\end{aligned}
\end{equation}
Here and below we will be parametrising the S~matrix in terms of the Zhukovsky variables~$x^\pm$, rather than of the momentum~$p$ (or the rapidity~$u$). While the Zhukovsky variables are constrained (unlike~$p$ or~$u$ which are unconstrained), they are quite useful because the expressions which we will encounter will be rational when expressed in the Zhukovsky variables.

\paragraph{Left-left block.}
Using the above equations it is possible to determine the various blocks of the S~matrix. For instance, when both particles are in the $(\phi^{\B}_{\L}|\varphi^{\F}_{\L})$ representation, which is the case in the string (as opposed to mirror) theory, we have
\begin{equation}
\label{eq:llblock}
\begin{aligned}
    \gen{S}\,|\phi_{\L,1}^\B\phi_{\L,2}^\B\rangle &= A^{\L\L}_{12}\,|\phi_{\L,2}^\B\phi_{\L,1}^\B\rangle,&\quad
    \gen{S}|\phi_{\L,1}^\B\varphi_{\L,2}^\F\rangle&= B^{\L\L}_{12}|\varphi_{\L,2}^\F\phi_{\L,1}^\B\rangle + C^{\L\L}_{12}|\phi_{\L,2}^\B\varphi_{\L,1}^\F\rangle,\\
    \gen{S}\,|\varphi_{\L,1}^\F\varphi_{\L,2}^\F\rangle &= F^{\L\L}_{12}\,|\varphi_{\L,2}^\F\varphi_{\L,1}^\F\rangle,&\quad
    \gen{S}\,|\varphi_{\L,1}^\F\phi_{\L,2}^\B\rangle&= D^{\L\L}_{12}|\phi_{\L,2}^\B\varphi_{\L,1}^\F\rangle + E^{\L\L}_{12}|\varphi_{\L,2}^\F\phi_{\L,1}^\B\rangle,
\end{aligned}
\end{equation}
where the six coefficients on the right-hand side can be determined up to an overall normalisation, see for instance~\cite{Eden:2021xhe} for a summary in this notation.

\paragraph{Left-right block.} In a similar way, we can determine the other blocks. For instance, the left-right block, when the left representation is as above and the right is $(\phi^{\F}_{\R}|\varphi^{\B}_{\R})$ (again, like it is the case in string theory), we have
\begin{equation}
\label{eq:lrblock}
\begin{aligned}
    \gen{S}\,|\phi_{\L,1}^\B\varphi_{\R,2}^\B\rangle&= A^{\L\R}_{12}|\varphi_{\R,2}^\B\phi_{\L,1}^\B\rangle + B^{\L\R}_{12}|\phi_{\R,2}^\F\varphi_{\L,1}^\F\rangle,
    &\quad
    \gen{S}\,|\phi_{\L,1}^\B\phi_{\R,2}^\F\rangle &= C^{\L\R}_{12}\,|\phi_{\R,2}^\F\phi_{\L,1}^\B\rangle,\\
    \gen{S}|\varphi_{\L,1}^\F\phi_{\R,2}^\F\rangle&= E^{\L\R}_{12}|\phi_{\R,2}^\F\varphi_{\L,1}^\F\rangle + F^{\L\R}_{12}|\varphi_{\R,2}^\B\phi_{\L,1}^\B\rangle,&\quad
    \gen{S}\,|\varphi_{\L,1}^\F\varphi_{\R,2}^\B\rangle &= D^{\L\R}_{12}\,|\varphi_{\R,2}^\B\varphi_{\L,1}^\F\rangle.
\end{aligned}
\end{equation}

\paragraph{Other blocks.} The other blocks, including those involving massless particles, can be found \textit{e.g.} in~\cite{Eden:2021xhe}, and we will not report them here.

\paragraph{Highest- and lowest-weight states in the mirror theory.} Suppose that we have worked out the S~matrix for the string model, as indeed it has been done in~\cite{Borsato:2014hja}, see also~\cite{Eden:2021xhe} for a concise illustration of the results. We now want to work out the S~matrix of the mirror model. Following~\cite{Arutyunov:2007tc}, see also~\cite{Frolov:2021zyc}, we know that we may analytically continue the Zhukovsky variables to do so. However, we have seen that the grading of highest-/lowest-weight states is reversed with respect to the string model. Effectively, in the two-particle representation~\eqref{eq:twoparticlerepr}, we want to reverse the highest- and lowest-weight state. In other words, a representation of the form (highest-weight$|$lowest-weight) $(\phi^\B|\varphi^\F)$ takes the form $(\phi^\F|\varphi^\B)$. This does not mean that the statistics of the particles is flipped, but rather that the highest- and lowest-weight states are reversed, $\phi^\B\leftrightarrow\varphi^\B$ and $\varphi^\F\leftrightarrow\phi^\F$.
In practice, because our model involves both left- and right-representations in the string model, which have opposite highest-weight states, going from the string to the mirror model amounts to exchanging the left- and right representation.

\paragraph{Full S matrix.}
The full S~matrix scatters four-dimensional irreducible representations of $\psu(1|1)^{\oplus4}$ c.e., rather than two-dimensional representations of $\psu(1|1)^{\oplus2}$~c.e.\ as described above.%
\footnote{%
The $\psu(1|1)^{\oplus2}$~c.e.\ S~matrix is also physically interesting in and of itself, as it yields the worldsheet S matrix of the~$AdS_3\times S^3\times S^3\times S^1$ background~\cite{Borsato:2012ud,Borsato:2015mma}.
}
Due to the factorised structure described in section~\ref{sec:particles:algebra}, we have that the full S~matrix can be written block by block, up to a pre-factor, in each block, as a graded tensor product
\begin{equation}
\label{eq:Stensor}
    \gen{S}_{\LL}(x^\pm,y^\pm) \,\hat{\otimes}\, \gen{S}_{\LL}(x^\pm,y^\pm)\,,\qquad
    \gen{S}_{\LR}(x^\pm,y^\pm) \,\hat{\otimes}\, \gen{S}_{\LR}(x^\pm,y^\pm)\,,\qquad\dots\,,
\end{equation}
where $\hat{\otimes}$ denotes the graded tensor product: writing the indices as $A=aa'$, $\dots$ $D=dd'$, we have
\begin{equation}
    (S \hat{\otimes} S')_{AB}^{CD} = (-1)^{F_{a'}F_{b}+F_{c'}F_{d}}\  S_{ab}^{cd}\  {S'}_{a'b'}^{c'd'}\,,
\end{equation}
where $F$ indicates the Fermion number of the various excitations.
Once again, going from the string to the mirror model is that the notion of which particles are highest-weight states and which ones are lowest-weight states is reversed with respect to the string model. For instance, if on the string model we have as highest-weight states $Y\sim\phi_\L^\B\otimes\phi_\L^\B$ and $\tilde{Z}\sim\phi_\R^\F\otimes\phi_\R^\F$, in the mirror model it is natural to have instead $\tilde{Y}\sim\phi_\L^\F\otimes\phi_\L^\F$ and $Z\sim\phi_\R^\F\otimes\phi_\R^\B$. Once again, a neat way to account for this difference is to say that \textit{the left- and right- representations are swapped when going from the string to the mirror model}. (In principle, the same analysis applies to the massless representations too, but they are actually indistinguishable as far as one can tell from looking at $\psu(1|1)^{\oplus4}$ centrally extended.)
Of course since the notion of the highest-weight state is a matter of convention (and besides, the model is left-right symmetric), we are always allowed to redefine what we mean by left- and right.

%\paragraph{A comment on $AdS_3\times S^3\times S^3\times S^1$.}
%We conclude noting that this difference is likely to be more relevant in the case of~$AdS_3\times S^3\times S^3\times S^1$~\cite{Borsato:2012ud,Borsato:2015mma}. For that model, we do not have the tensor-product structure~\eqref{eq:Stensor}. Instead, it is built directly out of the $4\times4$ S matrix $\gen{S}(p_1,p_2)$ introduced above. As a result, in that case it will be important to keep track of the signs due to the choice of symmetric or antisymmetric bound states.

\section{Algebraic Bethe Ansatz, factorised} 
\label{sec:BAonecopy}

We consider the factorised S matrix which encodes the scattering of the following four $2$-dimensional representations
\begin{equation}
\label{eq:fourirreps}
\rho_\L = ( \phi_\L^\B | \varphi_\L^\F)\,, \qquad \rho_\R = (\phi_\R^\F | \varphi_\R^\B)\,, \qquad \rho_\o = ( \phi_\circ^\B | \varphi_\circ^\F)\,, \qquad \rho_\op = (\phi_\circ^\F | \varphi_\circ^\B)\,.
\end{equation} 
Here we have made a definite choice for the statistics of left and right representations (denoted by B and F for Bosons and Fermions, respectively), which is the correct choice in the string model. For massless modes we find it convenient to introduce both possibilities for the statistics (because both will eventually be needed to construct the complete S~matrix of the string model). This will allow us to explicitly keep track of the relevant Fermion signs. As discussed in section~\ref{sec:particles:signs}, it will be easy to obtain the mirror S~matrix (and eventually transfer matrix) by swapping the notion of left- and right-representations.  Note that both $\o$ and $\op$ are massless. When the precise nature of the massless particle is not relevant%
\footnote{%
In particular for the S matrix elements, \textit{e.g.}~when the second particle is massless, $A^\LL_{12}\rightarrow  A^\LO_{12} = A^\Lo_{12}=A^\Lop_{12}$, where the second Zhukovsky variable obeys $x_2^+=1/x_2^-$.
} 
we will use the unifying notation $\O = \{\o,\op\}$. 

The only subtlety in deriving the Bethe equations is that we are dealing with several distinct irreducible representations for fundamental particles, which is a little unusual.
Restricting to the four representations in~\eqref{eq:fourirreps}, we have $8$ single-particle states, and hence $64$ states in $4\times4=16$ four-dimensional two-particle representations. Hence the S~matrix will be a $64 \times 64$ matrix with a priori $256$ blocks. But due to symmetries, most of these blocks vanish. For instance, if one has two incoming particles belonging respectively to $\rho_\L$ and $\rho_\o$ then so are the two outgoing particles, there cannot be the production of a  state in $\rho_\R$ or $\rho_\op$. Moreover, there is no reflection amplitude so that (in the notation where the S~matrix permutes the elements of the tensor product) $\rho_\L\otimes\rho_\o$ yields precisely~$ \rho_\o\otimes\rho_\L$. Hence, we only have $16$ non-trivial blocks (all $4 \times 4$ matrices). Each of these blocks has only six non-vanishing entries, as it can be seen \textit{e.g.}\ from~\eqref{eq:llblock} and~\eqref{eq:lrblock}.
Bearing all this in mind the algebraic Bethe ansatz can be worked out having care of keeping track of the various blocks.

We consider $K$ physical particles with momenta $p_1, \dots, p_K$ (Zhukovsky variables $x^\pm_j$, $j=1,\dots,K$) in the physical Hilbert space $H=H_1 \otimes \dots \otimes H_K$ as well as one unphysical particle with momentum $p$ (Zhukovsky variables $x^\pm$) in the auxiliary space $H_a = \amsmathbb{C}^2 \oplus \amsmathbb{C}^2 \oplus \amsmathbb{C}^2 \oplus \amsmathbb{C}^2$. Scattering this unphysical particle through all the physical ones gives rise to the monodromy matrix
\begin{equation}
    \Mono(x^\pm) = \gen{S}_{aK}(x^\pm,x^\pm_K) \dots \gen{S}_{a1}(x^\pm,x^\pm_1)\,,
\end{equation}
where the subscripts indicate on which spaces the two-body S matrices act. Notice that $\Mono(x^\pm)$ also depends on the variables $\{x^\pm_j\}$ but it will be convenient to simplify notation and omit this explicit dependence.
In auxiliary space, the monodromy is a block-diagonal matrix with four blocks of $2 \times 2$ dimensional matrices,
\begin{equation} \begin{aligned}
\Mono(x^\pm) &= \text{diag}\Big(\Mono_\L(x^\pm),\, \Mono_\R(x^\pm),\, \Mono_\o(x^\pm),\, \Mono_\op(x^\pm) \Big), \\
\Mono_\star(x^\pm) &= \begin{pmatrix}
\MA_\star(x^\pm) && \MB_\star(x^\pm) \\
\MC_\star(x^\pm) && \MD_\star(x^\pm) 
\end{pmatrix},\qquad  \star \in \{\L,\R,\o,\op \}\,.
\end{aligned}
\end{equation}
 The entries are themselves operators acting on the physical space (and which could hence be represented as $2^K \times 2^K$ matrices). 
For each representation we define the corresponding transfer matrix, obtained by taking the supertrace over the auxiliary space, $\Tran_\star(x^\pm) = \text{STr}_a \Mono_\star(x^\pm)$. Taking into account the grading this gives
\begin{align}
\Tran_\L(x^\pm) &=+ \MA_\L(x^\pm) - \MD_\L(x^\pm)\,, \qquad
&\Tran_\R(x^\pm) &=- \MA_\R(x^\pm) + \MD_\R(x^\pm), \\
\Tran_\o(x^\pm) &= +\MA_\o(x^\pm) - \MD_\o(x^\pm)\,, \qquad
&\Tran_\op(x^\pm) &= - \MA_\op(x^\pm) + \MD_\op(x^\pm)\,.
\end{align}
These are operators on the physical space.
\subsection{Commutation relation from RTT relations}
The monodromy matrix satisfies the RTT relation
\begin{equation}
    \gen{S}(x^\pm,y^\pm) \ \Mono(x^\pm) \check{\otimes} \Mono(y^\pm) = \Mono(y^\pm) \check{\otimes} \Mono(x^\pm) \ \gen{S}(x^\pm, y^\pm)\,,
\end{equation}
where $\check{\otimes}$ is defined as follows: for two matrices $A_a{}^c$ and $B_b{}^d$ one has
\begin{equation}
    (A \check{\otimes} B)_{ab}^{cd} = (-1)^{F_b(F_a+F_c)} A_a{}^c B_b{}^d~,
\end{equation}
with $F_j=0$ for Bosonic excitations and $F_j=1$ for Fermionic ones.
Here we summarise the important commutation relation among the operators acting in physical space. For the diagonal elements we have
\begin{equation}
\com{\MA_{\star}(x^\pm) }{\MA_{\star'}(y^\pm)} = \com{\MD_{\star}(x^\pm) }{\MD_{\star'}(y^\pm)} =  0\,,
\end{equation}
as well as, for instance,~\footnote{To simplify notation we have suppressed the arguments of the S matrix elements appearing on the right-hand side of the various commutation relations. For instance in \eqref{eq:RTTAD} one has $C^\LL = C^\LL(x^\pm,y^\pm)$ as well as $D^\LL = D^\LL(x^\pm,y^\pm)$.}
\begin{align}
\label{eq:RTTAD}
\com{\MA_\L(x^\pm)}{\MD_\L(y^\pm)} &= -\frac{C^\LL}{D^\LL} \Big(\MB_\L(x^\pm) \MC_\L(y^\pm)-\MB_\L(y^\pm) \MC_\L(x^\pm)\Big)\,, \\
\com{\MA_\L(x^\pm)}{\MD_\R(y^\pm)} &= -\frac{B^\LR}{E^\LR} \Big(\MB_\L(x^\pm) \MC_\R(y^\pm)-\MB_\R(y^\pm) \MC_\L(x^\pm)\Big)\,, \\
\com{\MA_\L(x^\pm)}{\MD_\o(y^\pm)} &= -\frac{C^\LO}{D^\LO} \Big(\MB_\L(x^\pm) \MC_\o(y^\pm)-\MB_\o(y^\pm) \MC_\L(x^\pm)\Big)\,, \\
\com{\MA_\L(x^\pm)}{\MD_\op(y^\pm)} &= +\frac{C^\LO}{D^\LO} \Big(\MB_\L(x^\pm) \MC_\op(y^\pm)-\MB_\op(y^\pm) \MC_\L(x^\pm)\Big)\,.
\end{align}
The commutation relations involving a diagonal element and a raising operator are 
\begin{align} \label{eq:RTTLL}
\MA_\L(x^\pm)\, \MB_\L(y^\pm) &= -\frac{F^\LL}{D^\LL} \MB_\L(y^\pm)\, \MA_\L(x^\pm)+ \frac{C^\LL}{D^\LL} \MB_\L(x^\pm)\, \MA_\L(y^\pm)\,, \\ \label{eq:RTTLR}
\MA_\L(x^\pm)\, \MB_\R(y^\pm) &= -\frac{D^\LR}{E^\LR} \MB_\R(y^\pm) \MA_\L(x^\pm) -\frac{B^\LR}{E^\LR} \MB_\L(x^\pm)\, \MA_\R(y^\pm)\,, \\ \label{eq:RTTLo}
\MA_\L(x^\pm)\, \MB_\o(y^\pm) &= -\frac{F^\LO}{D^\LO}  \MB_\o(y^\pm)\, \MA_\L(x^\pm) +\frac{C^\LO}{D^\LO} \MB_\L(x^\pm)\, \MA_\o(y^\pm)\,, \\ \label{eq:RTTLop}
\MA_\L(x^\pm)\, \MB_\op(y^\pm) &= -\frac{F^\LO}{D^\LO}  \MB_\op(y^\pm)\, \MA_\L(x^\pm) -\frac{C^\LO}{D^\LO} \MB_\L(x^\pm)\, \MA_\op(y^\pm)\,.
\end{align}
The same equations hold when replacing $\MA$ with~$\MD$. Looking at the right representation, we have
\begin{align}
\MA_\R(x^\pm)\, \MB_\L(y^\pm) &= \frac{C^\RL}{A^\RL} \MB_\L(y^\pm)\, \MA_\R(x^\pm)+ \frac{F^\RL}{A^\RL} \MB_\R(x^\pm)\, \MA_\L(y^\pm)\,, \\
\MA_\R(x^\pm)\, \MB_\R(y^\pm) &= \frac{A^\RR}{B^\RR} \MB_\R(y^\pm)\, \MA_\R(x^\pm) -\frac{E^\RR}{B^\RR} \MB_\R(x^\pm)\, \MA_\R(y^\pm)\,, \\
\MA_\R(x^\pm)\, \MB_\o(y^\pm) &= \frac{C^\RO}{A^\RO}  \MB_\o(y^\pm)\, \MA_\R(x^\pm) +\frac{F^\RO}{A^\RO} \MB_\R(x^\pm)\, \MA_\o(y^\pm)\,, \\
\MA_\R(x^\pm)\, \MB_\op(y^\pm) &= \frac{C^\RO}{A^\RO}  \MB_\op(y^\pm)\, \MA_\R(x^\pm) -\frac{F^\RO}{A^\RO} \MB_\R(x^\pm)\, \MA_\op(y^\pm)\,.
\end{align}
To obtain the commutation relations involving $\MA_\o(x^\pm)$ and $\MD_\o(x^\pm)$ one can just appropriately replace~``L'' with ``$\o$'' in \eqref{eq:RTTLL}--\eqref{eq:RTTLop}. To obtain the commutation relations involving $\MA_\op(x^\pm)$ and $\MD_\op(x^\pm)$ one needs to replace~``L'' with ``$\op$'' but some signs will be different due to the fact that the highest weight state is fermionic, leading to
\begin{align}
\MA_\op(x^\pm)\, \MB_\L(y^\pm) &= -\frac{F^\OL}{D^\OL} \MB_\L(y^\pm)\, \MA_\op(x^\pm)-\frac{C^\OL}{D^\OL} \MB_\op(x^\pm)\, \MA_\L(y^\pm)\,, \\
\MA_\op(x^\pm)\, \MB_\R(y^\pm) &= -\frac{D^\OR}{E^\OR} \MB_\R(y^\pm)\, \MA_\op(x^\pm) +\frac{B^\OR}{E^\OR} \MB_\op(x^\pm)\, \MA_\R(y^\pm)\,, \\
\MA_\op(x^\pm)\, \MB_\o(y^\pm) &= -\frac{F^\OO}{D^\OO}  \MB_\o(y^\pm)\, \MA_\op(x^\pm) -\frac{C^\OO}{D^\OO} \MB_\op(x^\pm) \MA_\o(y^\pm)\,, \\
\MA_\op(x^\pm)\, \MB_\op(y^\pm) &= -\frac{F^\OO}{D^\OO}  \MB_\op(y^\pm)\, \MA_\op(x^\pm) +\frac{C^\OO}{D^\OO} \MB_\op(x^\pm)\, \MA_\op(y^\pm)\,.
\end{align}
Other important commutation relations involve two creation operators, like
\begin{align}
\label{eq:BBi}
\MB_\L(x^\pm)\, \MB_\L(y^\pm) &= +\frac{F^\LL}{A^\LL}  \MB_\L(y^\pm)\, \MB_\L(x^\pm)\,, \\
\MB_\L(x^\pm)\, \MB_\R(y^\pm) &= -\frac{D^\LR}{C^\LR}  \MB_\R(y^\pm)\, \MB_\L(x^\pm)\,, \\
\MB_\L(x^\pm)\, \MB_\o(y^\pm) &= +\frac{F^\LO}{A^\LO} \MB_\o(y^\pm)\, \MB_\L(x^\pm)\,, \\
\MB_\L(x^\pm)\, \MB_\op(y^\pm) &= +\frac{F^\LO}{A^\LO}  \MB_\op(y^\pm)\, \MB_\L(x^\pm)\,, 
\end{align}
and
\begin{align}
\MB_\R(x^\pm)\, \MB_\R(y^\pm) &= +\frac{A^\RR}{F^\RR}  \MB_\R(y^\pm)\, \,\MB_\R(x^\pm)\,, \\
\MB_\R(x^\pm)\, \MB_\o(y^\pm) &= -\frac{C^\RO}{D^\RO}  \MB_\o(y^\pm)\, \MB_\R(x^\pm)\,, \\
\MB_\R(x^\pm)\, \MB_\op(y^\pm) &= -\frac{C^\RO}{D^\RO}  \MB_\op(y^\pm)\, \MB_\R(x^\pm)\,.
\label{eq:BBf}
\end{align}
We note that, while two $\MB_\star$ operators do not commute, they do give rise to the same state up to a scalar coefficient.

It is natural to wonder whether using different $\MB$ operators results in genuinely different states. This is important to correctly count the number of states in the theory. In ref.~\cite{Borsato:2012ss} it was argued from the coordinate Bethe ansatz that there is only one type of auxiliary root valid both for left and right representations (as well as massless, though that was only discussed in~\cite{Borsato:2014hja}), which would mean that the action of $\mathcal{B}_{\L}(x^\pm)$ should be equivalent to that of $\mathcal{B}_{\R}(y^\pm)$ provided that $x^\pm$ and $y^\pm$ are related in some specific way.
This is actually the case. In particular, in appendix~\ref{app:Null} we show, using the above RTT relations, that the action of two creation operators~$\MB_\star$ in different representations with suitably related rapidities is equivalent.

%{\bf \color{red} I actually think that those should give minus one, not one. (repeated rapidities should always give zero.) it probably depends on whether you used the permutation or graded permutation to define the R-matrix.} {\color{red} If I use standard RTT relation 
%\begin{equation}
%    R(p,q) (M(p) \otimes 1) (1 \otimes M(q)) = (1 \otimes M(q))(M(p) \otimes 1) R(p,q) 
%\end{equation}
%with $R$ obtained from the $S$ of this file through the graded permutation operator, $R=\Pi^g S$ then I get a minus sign, but I am confused why having same rapidity should give a $-1$ prefactor. For XXX spin chain the creation operator satisfies $B(u)B(v)=B(v)B(u)$ for all $u,v$. Is the difference that $B(u)$ in the XXX spin chain corresponds to bosonic operator while here it is fermionic? Also, my commutation relations, including this $+1$ factor, match the result of 2103.16972 p.24.}

Finally, note that in simplifying the RTT relations we used the following identities
\begin{align}
A^\LL F^\LL + B^\LL D^\LL - C^\LL E^\LL &= 0\,, \\
A^\RR F^\RR + B^\RR D^\RR - C^\RR E^\RR &= 0\,, \\
A^\LR E^\LR - B^\LR F^\LR + C^\LR D^\LR &=0\,, \\
A^\RL E^\RL - B^\RL F^\RL + C^\RL D^\RL &= 0\,.
\end{align}
More identities follow by taking the massless limit $\L \rightarrow \O$.
\subsection{Vacuum}
%For the choice of vacuum, all four highest weights 
%\begin{equation}
%\ket{0} = \ket{\phi_\L^\B}^{\otimes L}~, \qquad \ket{0} = \ket{\phi_\R^\F}^{\otimes L}~, \qquad \ket{0} = \ket{\phi_\o^\B}^{\otimes L}~, \qquad \ket{0} = \ket{\phi_\op^\F}^{\otimes L}
%\end{equation}
%are such that
%\begin{equation}
%(T_\star)_1{}^1 \ket{0} =\Omega_{\star,1} \ket{0}~, \qquad (T_\star)_2{}^2 \ket{0} =\Omega_{\star,2} \ket{0}~, \qquad (T_\star)_2{}^1 \ket{0} =0~.
%\end{equation}
We define the vacuum as a tensor product of highest-weight states chosen from all the possible representations. Setting $K_\o+K_\op=K_\O$ we write (for massless particles $x=x^+=1/x^-$)
\begin{equation}
\ket{0} = \bigotimes_{j=1}^{K_\L}\ket{\phi_\L^\B(x_{\L,j}^\pm)} \bigotimes_{j=1}^{K_\R} \ket{\phi_\R^\F(x_{\R,j}^\pm)} \bigotimes_{j=1}^{K_\o} \ket{\phi_\O^\B(x_{j})}\bigotimes_{j=K_\o+1}^{K_\O} \ket{\phi_\O^\F(x_{j})}~, 
\end{equation}
so that as desired one has that the four ``lowering'' entries~$\MC_\star$ in each block of the transfer matrix annihilate the vacuum,
\begin{equation}
    \MC_\star(y^\pm) \ket{0} =0\,.
\end{equation}
The diagonal entries act diagonally,
\begin{equation}
\MA_\star(y^\pm)\ket{0} =\Omega_{\star,1}(y^\pm,\{x^\pm_j\}) \ket{0}\,, \qquad \MD_\star(y^\pm) \ket{0} =\Omega_{\star,2}(y^\pm,\{x^\pm_j\}) \ket{0}\,.
\end{equation}
We will often keep the dependence on the various $\{x_j^\pm\}$ implicit. In the above
\begin{equation} \begin{aligned}
\Omega_{\L,1}(y^\pm) &= \prod_{j=1}^{K_\L} A^\LL(y^\pm,x_{\L,j}^\pm)\prod_{j=1}^{K_\R} C^\LR(y^\pm,x_{\R,j}^\pm)\prod_{j=1}^{K_\O} A^\LO(y^\pm,x_{j})~,\\
\Omega_{\L,2}(y^\pm) &= \prod_{j=1}^{K_\L} D^\LL(y^\pm,x_{\L,j}^\pm)\prod_{j=1}^{K_\R} -E^\LR(y^\pm,x_{\R,j}^\pm)\prod_{j=1}^{K_\O} D^\LO(y^\pm,x_{j})~, \\
\Omega_{\R,1}(y^\pm) &= \prod_{j=1}^{K_\L} D^\RL(y^\pm,x_{\L,j}^\pm)\prod_{j=1}^{K_\R} -F^\RR(y^\pm,x_{\R,j}^\pm)\prod_{j=1}^{K_\O} D^\RO(y^\pm,x_{j})~, \\
\Omega_{\R,2}(y^\pm) &= \prod_{j=1}^{K_\L} A^\RL(y^\pm,x_{\L,j}^\pm)\prod_{j=1}^{K_\R}B^\RR(y^\pm,x_{\R,j}^\pm)\prod_{j=1}^{K_\O} A^\RO(y^\pm,x_{j})~,
\end{aligned}
\end{equation}
and 
\begin{equation} \begin{aligned}
\Omega_{\O,1}(y^\pm) \equiv \Omega_{\o,1}(y^\pm) =\Omega_{\op,1}(y^\pm) &= \prod_{j=1}^{K_\L} A^\OL(y^\pm,x_{\L,j}^\pm)\prod_{j=1}^{K_\R} C^\OR(y^\pm,x_{\R,j}^\pm)\prod_{j=1}^{K_\O} A^\OO(y^\pm,x_{j})~, \\
\Omega_{\O,2}(y^\pm) \equiv  \Omega_{\o,2}(y^\pm)=\Omega_{\op,2}(y^\pm) &= \prod_{j=1}^{K_\L} D^\OL(y^\pm,x_{\L,j}^\pm)\prod_{j=1}^{K_\R} -E^\OR(y^\pm,x_{\R,j}^\pm)\prod_{j=1}^{K_\O} D^\OO(y^\pm,x_{j})~.
\end{aligned}
\end{equation}

The remaining entries, $\MB_\star$, correspond to raising operators that yield new states. We will see in the next section how to act with these various operators.

\subsection{One excitation above the vacuum}
If we want to create a single excitation over the vacuum, we have four choices, corresponding to picking our representation label~$\star$ to be $\star \in \{\text{L},\text{R},\o,\op \}$. Hence we may write such a generic state as
\begin{equation}
\ket{\Phi_\star (y^\pm)} =  \MB_\star(y^\pm) \ket{0}\,.
\end{equation}
We may now use the commutation relations for the entries of the monodromy matrix to  diagonalise the transfer matrix.

\paragraph{Case $\star = \text{L}$.}
Using the commutation relation we deduce that
\begin{align}
\label{eq:eigenvalueeqL}
\Tran_\L (x^\pm) \ket{\Phi_\L (y^\pm)} &=   \Lambda_{\L}(x^\pm,y^-)\, \ket{\Phi_\L (y^\pm)}  + \Delta_{\L}(y^\pm)\, Z_\L(x^\pm)  \ket{0}, \\
\Tran_\R (x^\pm) \ket{\Phi_\L (y^\pm)} &= \Lambda_{\R}(x^\pm,y^-) \, \ket{\Phi_\L (y^\pm)}  + \Delta_{\L}(y^\pm)\, Z_\R(x^\pm)  \ket{0}, \\
\Tran_\o (x^\pm) \ket{\Phi_\L (y^\pm)} &=   \Lambda_{\o}(x^\pm,y^-)\, \ket{\Phi_\L (y^\pm)}  + \Delta_{\L}(y^\pm) \, Z_\o(x^\pm)  \ket{0}, \\
\Tran_\op (x^\pm)\ket{\Phi_\L (y^\pm)} &=   \Lambda_{\op}(x^\pm,y^-)\,\ket{\Phi_\L (y^\pm)}  + \Delta_{\L}(y^\pm)\, Z_\op(x^\pm)  \ket{0}, 
\label{eq:eigenvalueeqLend}
\end{align}
where we denoted by $\Delta_{\star}$ the difference of the vacuum eigenvalues (which implicitly also depends on all $\{x_j^\pm\}$)
\begin{equation}
    \Delta_{\star}(y^\pm) = \Omega_{\star,1}(y^\pm) - \Omega_{\star,2}(y^\pm)\,,
\end{equation}
and we indicated the eigenvalues of the transfer matrix by
\begin{align}
\Lambda_\L(x^\pm,y^-) &= - \frac{F^\LL(x^\pm,y^\pm)}{D^\LL(x^\pm,y^\pm)}\, \Delta_\L(x^\pm)&= \sqrt{\frac{x^-}{x^+}}\frac{y^--x^+}{y^--x^-}&\,\Delta_\L(x^\pm)\,,\\
\Lambda_\R(x^\pm,y^-) &= - \frac{C^\RL(x^\pm,y^\pm)}{A^\RL(x^\pm,y^\pm)}\, \Delta_\R(x^\pm)&=-\sqrt{\frac{x^+}{x^-}} \frac{1-y^- x^-}{1-y^- x^+}&\,\Delta_\R(x^\pm)\,, \\
\Lambda_\o(x^\pm,y^-) &= - \frac{F^\OL(x^\pm,y^\pm)}{D^\OL(x^\pm,y^\pm)}\, \Delta_\O(x^\pm)&=\sqrt{\frac{x^-}{x^+}}\frac{y^--x^+}{y^--x^-}&\,\Delta_\O(x^\pm)\,,\\
\Lambda_\op(x^\pm,y^-) &= + \frac{F^\OL(x^\pm,y^\pm)}{D^\OL(x^\pm,y^\pm)}\, \Delta_\O(x^\pm)&=-\sqrt{\frac{x^-}{x^+}}\frac{y^--x^+}{y^--x^-}&\,\Delta_\O(x^\pm)\,.
\end{align}
Note that the eigenvalues depend only on~$y^-$. We also introduced the short-hands
\begin{equation} \begin{aligned}
Z_\L(x^\pm) &=  \frac{C^\LL(x^\pm,y^\pm)}{D^\LL(x^\pm,y^\pm)}  \MB_\L(x^\pm)\,, &\qquad Z_\R(y^\pm) &= - \frac{F^\RL(x^\pm,y^\pm)}{A^\RL(x^\pm,y^\pm)} \MB_\R(x^\pm)\,, \\
Z_\o(x^\pm) &= \frac{C^\OL(x^\pm,y^\pm)}{D^\OL(x^\pm,y^\pm)} \MB_\o(x^\pm)\,, &\qquad Z_\op(x^\pm) &= \frac{C^\OL(x^\pm,y^\pm)}{D^\OL(x^\pm,y^\pm)} \MB_\op(x^\pm)\,.
\end{aligned}
\end{equation}

For $\ket{\Phi_\L (y^\pm)} $ to be an eigenstate of the transfer matrices we need the second term in the right-hand-side of each line of~\eqref{eq:eigenvalueeqL}--\eqref{eq:eigenvalueeqLend}  to vanish.
All the four conditions can be solved by a single equation, which is a constraint on~$y$, namely
% \begin{equation}
% \label{eq:Bethe1L}
% \Delta_{\L}(y^\pm)=0\,,\qquad\Leftrightarrow\qquad
% \frac{\Omega_{\L,1}}{\Omega_{\L,2}}(y^-) =1\,.
% \end{equation}
\begin{equation}
\label{eq:Bethe1L}
\Delta_{\L}(y^\pm)=0\qquad\Leftrightarrow\qquad \left(\frac{\Omega_{\L,1}}{\Omega_{\L,2}}\right)\big(y^-,\{x_j^\pm\}\big)=1\,,
\end{equation}
which is independent from~$y^+$.
The last equation takes the explicit form of an equation for the dummy variable~$y^-$
\begin{equation}
\label{eq:BetheLeft}
    \prod_{\J\in \{\L,\R,\O\}}
\prod_{j=1}^{K_\J} S^\mathrm{I,II}_{\J\L} (x_{\J,j}^\pm,y^-) =1\,,
\end{equation}
where we introduced the auxiliary S~matrices
\begin{align}
S^\mathrm{I,II}_\LL (x^\pm,y) &= S^\mathrm{I,II}_\OL (x^\pm,y) =  \sqrt{\frac{x^-}{x^+}}\frac{y-x^+}{y-x^-}\,, \\
S^\mathrm{I,II}_\RL (x^\pm,y) &=  \sqrt{\frac{x^+}{x^-}} \frac{1-y x^-}{1-y x^+}\,.
\end{align}

\paragraph{Regularity of the transfer matrix.}
Let us briefly comment on an equivalent way to derive the Bethe equation~\eqref{eq:BetheLeft}.
Notice that $D^\LL(x^\pm,y^\pm)$ has a zero when $x^-=y^-$. Requiring that the residue vanishes is equivalent to imposing the Bethe equation \eqref{eq:Bethe1L}. The same conclusion follows from observing that $A^\RL(x^\pm,y^\pm)$ has a zero at $x^+=1/y^-$. This gives
\begin{equation}
    0=\Delta_\R(x^\pm)\Big|_{x^+=1/y^-}\,,
\end{equation}
which is the case if and only if~\eqref{eq:Bethe1L}~holds.
Similarly, we need to ensure that the massless eigenvalues are regular. This means that the zero in $D^\OL(x^\pm,y^\pm)$, which is again at $x^-=y^-$, is compensated by a zero in $\Delta_\O(x^\pm)$ evaluated at $x^-=y^-$. This is not an issue because these expressions have the same form as in the ``left'' case, up to imposing the massless condition $x^+=1/x^-$, which does not affect the discussion but does simplify the form of the auxiliary S~matrices $S^{\mathrm{I,II}}$.

\paragraph{On the auxiliary variables.}
Both the Bethe equations and the eiganvalues of the transfer matrix depend on the auxiliary variable~$y^\pm$ only through $y^-$. Moreover, $y^-$ itself is a dummy variable which does not affect the momentum or energy and is determined by the Bethe equations. Hence it is convenient to just rename it as~$y$. We refrain from doing so for the moment, to make the subsequent comparison with the ``right'' equations clearer.

\paragraph{Case $\star = \text{R}$.}
The derivation is completely analogous to the previous case and we will not repeat it. The eigenvalues only depend on $y^+$ and are
\begin{align}
\Lambda_\L(x^\pm,y^+) &= - \frac{D^\LR(x^\pm,y^\pm) }{E^\LR(x^\pm,y^\pm) }\,  \Delta_\L(x^\pm)&=\sqrt{\frac{x^-}{x^+}}\frac{\frac{1}{y^+}-x^+}{\frac{1}{y^+}-x^-}&\,  \Delta_\L(x^\pm)\,, \\
\Lambda_\R(x^\pm,y^+) &=  - \frac{A^\RR(x^\pm,y^\pm) }{B^\RR(x^\pm,y^\pm) }\,  \Delta_\R(x^\pm)&=- \sqrt{\frac{x^+}{x^-}} \frac{1-\frac{1}{y^+} x^-}{1-\frac{1}{y^+} x^+}&\,  \Delta_\R(x^\pm)\,, \\
\Lambda_\o(x^\pm,y^+) &= - \frac{D^\OR(x,y^\pm) }{E^\OR(x,y^\pm)}\,   \Delta_\O(x^\pm)&=\sqrt{\frac{x^-}{x^+}}\frac{\frac{1}{y^+}-x^+}{\frac{1}{y^+}-x^-}&\,  \Delta_\O(x^\pm)\,\,, \\
\Lambda_\op(x^\pm,y^+) &= + \frac{D^\OR(x,y^\pm) }{E^\OR(x,y^\pm)}\,   \Delta_\O(x^\pm)&=-\sqrt{\frac{x^-}{x^+}}\frac{\frac{1}{y^+}-x^+}{\frac{1}{y^+}-x^-}&\,  \Delta_\O(x^\pm)\,\,.
\end{align}

As it can be seen from the singularity of $B^\RR(x^\pm,y^\pm)$ at $x^+=y^+$, we need to impose the Bethe equation
\begin{equation}
\Delta_{\R}(y^\pm)=0\,,
\qquad\Leftrightarrow\qquad
\left(\frac{\Omega_{\R,1}}{\Omega_{\R,2}}\right)(y^+,\{x^\pm_j\}) =1\,.
\end{equation}
This gives rise to the equation for the dummhy variable~$y^+$
\begin{equation}
\prod_{\J \in \{\L,\R,\O\}}
\prod_{j=1}^{K_\J} S^\mathrm{I,II}_{\J\R} (x_{\J,j}^\pm,y^+) =1\,,
\end{equation}
with
\begin{align}
    S^\mathrm{I,II}_\LR (x^\pm,y) &= S^\mathrm{I,II}_\OR (x^\pm,y)  = \sqrt{\frac{x^-}{x^+}}\frac{\frac{1}{y}-x^+}{\frac{1}{y}-x^-}\,, \\
S^\mathrm{I,II}_\RR (x^\pm,y) &=  \sqrt{\frac{x^+}{x^-}} \frac{1-\frac{1}{y} x^-}{1-\frac{1}{y} x^+}\,.
\end{align}
This also guarantees that the apparent pole due to the zero of~$E^\LR(x^\pm,y^\pm)$ is absent (and similarly for the massless equations).

\paragraph{Case $\star = \O \in \{\o,\op\}$.}
In the massless case, the eigenvalues are 
\begin{align}
\Lambda_\L(x^\pm,y) &= - \frac{F^\LO(x^\pm,y) }{D^\LO(x^\pm,y) } \Delta_\L(x^\pm)&=\sqrt{\frac{x^-}{x^+}}\frac{y^--x^+}{y^--x^-}&\,\Delta_\L(x^\pm)\,,  \\
\Lambda_\R(x^\pm,y) &=  - \frac{C^\RO(x^\pm,y) }{A^\RO(x^\pm,y) }  \Delta_\R(x^\pm)&=-\sqrt{\frac{x^+}{x^-}} \frac{1-y^- x^-}{1-y^- x^+}&\,\Delta_\R(x^\pm)\,, \\
\Lambda_\o(x^\pm,y^\pm) &= - \frac{F^\OO(x^\pm,y^\pm) }{D^\OO(x^\pm,y^\pm)}  \Delta_\O(x^\pm)&=\sqrt{\frac{x^-}{x^+}}\frac{y^--x^+}{y^--x^-}&\,\Delta_\O(x^\pm)\,, \\
\Lambda_\op(x^\pm,y^\pm) &= + \frac{F^\OO(x^\pm,y^\pm) }{D^\OO(x^\pm,y^\pm)}  \Delta_\O(x^\pm)&=-\sqrt{\frac{x^-}{x^+}}\frac{y^--x^+}{y^--x^-}&\,\Delta_\O(x^\pm)\,,.
\end{align}
The Bethe equation is (as it can also be seen from the zero of $D^\OO(x,y)$)
\begin{equation}
\Delta_\O(y^\pm)=0\,,\qquad\Leftrightarrow\qquad
\left(\frac{\Omega_{\O,1}}{\Omega_{\O,2}}\right)(y^-) =1\,.
\end{equation}
This gives rise to the equation for the dummy variable~$y^-$
\begin{equation}
\label{eq:BetheMassless}
    \prod_{\J\in \{\L,\R,\O\}}
\prod_{j=1}^{K_\J} S^\mathrm{I,II}_{\J\O} (x_{\J,j}^\pm,y^-) =1
\end{equation}
with
\begin{align}
S^\mathrm{I,II}_\LO (x^\pm,y) &= S^\mathrm{I,II}_\OO (x^\pm,y) =  \sqrt{\frac{x^-}{x^+}}\frac{y-x^+}{y-x^-}\,, \\
S^\mathrm{I,II}_\RO (x^\pm,y) &=  \sqrt{\frac{x^+}{x^-}} \frac{1-y x^-}{1-y x^+}\,.
\end{align}
%\subsection{Two excitations above the vacuum}
%We make the following Ansatz 
%\begin{equation}
%\ket{\Phi_{\star_1 \star_2} (u_1,u_2)} =  (T_{\star_1})_1{}^2(u_1)  (T_{\star_2})_1{}^2(u_2)\ket{0}~.
%\end{equation}
%This is an eigenstate of  the transfer matrix provided one imposed the Bethe equations
%\begin{equation}
%\frac{\Omega_{\star_1,1}(u_1)}{\Omega_{\star_1,2}(u_1)} =1~, \qquad \frac{\Omega_{\star_2,1}(u_2)}{\Omega_{\star_2,2}(u_2)} =1~.
%\end{equation}
%The eigenvalues are
%\begin{align}
%\Lambda_\LL(y^\pm) &= \prod_{j=1}^2 - \frac{F^\LL(y^\pm,u_j)}{D^\LL(y^\pm,u_j)}\left(\Omega_{\L,1}(y^\pm) - \Omega_{\L,2}(y^\pm)\right)- \prod_{j=1}^2\frac{C^\RL(y^\pm,u_j)}{A^\RL(y^\pm,u_j)} \left(\Omega_{\R,1}(y^\pm) - \Omega_{\R,2}(y^\pm)\right)~,\\
%\Lambda_\RR(y^\pm) &= \prod_{j=1}^2 - \frac{D^\LR(y^\pm,u_j)}{E^\LR(y^\pm,u_j)}\left(\Omega_{\L,1}(y^\pm) - \Omega_{\L,2}(y^\pm)\right)- \prod_{j=1}^2\frac{A^\RR(y^\pm,u_j)}{B^\RR(y^\pm,u_j)} \left(\Omega_{\R,1}(y^\pm) - \Omega_{\R,2}(y^\pm)\right)~,\\
%\Lambda_\LR(y^\pm) &=  \frac{F^\LL(y^\pm,u_1)}{D^\LL(y^\pm,u_1)}\frac{D^\LR(y^\pm,u_2)}{E^\LR(y^\pm,u_2)}\left(\Omega_{\L,1}(y^\pm) - \Omega_{\L,2}(y^\pm)\right)- \frac{C^\RL(y^\pm,u_j)}{A^\RL(y^\pm,u_j)}\frac{A^\RR(y^\pm,u_j)}{B^\RR(y^\pm,u_j)} \left(\Omega_{\R,1}(y^\pm) - \Omega_{\R,2}(y^\pm)\right)~,
%\end{align}

\subsection{Several excitations above the vacuum}
For a state with several excitations above the vacuum, we make the Ansatz 
\begin{equation}
\label{eq:severalexfact}
\ket{\Phi(\vec{y})} =   \prod_{\star=\L,\R,\o, \op}\,\prod_{k=1}^{N_\star}\MB_\star(y^\pm_{\star,k}) \ket{0}\,.
\end{equation}
Here $N_\L,N_\R,N_\o,N_\op$ are the number of excitations. Since the operators do not commute, strictly speaking with the above short-hand we mean the \textit{ordered} product of operators (in the order in which the indices are written). However, as discussed around~\eqref{eq:BBi}--\eqref{eq:BBf}, reordering the creation operators gives the same state, as the commutation only yields a scalar prefactor.
The vector $\vec{y} = (y^\pm_{\L,1},\dots,y^\pm_{\L,N_\L},y^\pm_{\R,1},\dots,y^\pm_{\op,N_{\op}})$ gathers all the roots.
The first lower index, which labels the representation, is mostly useful to keep track of which root sits in which operator. Eventually we will have to solve for the rapidities~$y^\pm_{\star,j}$. We will discuss later how these types of roots are not really different.

The state \eqref{eq:severalexfact} is an eigenstate of the transfer matrices provided one imposes the Bethe equations
\begin{equation}
\Delta_\star(y_{\star,k}^\pm)=0~, \qquad k=1,\dots,N_\star~, \qquad \star \in \{\L,\R,\O\}~.
\end{equation}
where we found it convenient to also use the unifying notation $\O = \{ \o, \op\}$ for the auxiliary roots, with $N_\O = N_\o + N_\op$.
These can be written
% \begin{align}
% 1&=\prod_{j=1}^{L_\L} S^\mathrm{II,I}_\LL (y_k^\L,x_{j,\L}^\pm) \prod_{j=1}^{L_\R} S^\mathrm{II,I}_\LR(y_k^\L,x_{j,\R}^\pm)\prod_{j=1}^{L_\O} S^\mathrm{II,I}_\LO (y_k^\L,x_{j})~, \qquad k=1,\dots,N_\L~, \\
% 1&=\prod_{j=1}^{L_\L} S^\mathrm{II,I}_\RL (y_k^\R,x_{j,\L}^\pm) \prod_{j=1}^{L_\R} S^\mathrm{II,I}_\RR(y_k^\R,x_{j,\R}^\pm)\prod_{j=1}^{L_\O} S^\mathrm{II,I}_\RO (y_k^\R,x_{j})~, \qquad k=1,\dots,N_\R~, \\
% 1&=\prod_{j=1}^{L_\L} S^\mathrm{II,I}_\OL (y_k^\o,x_{j,\L}^\pm) \prod_{j=1}^{L_\R} S^\mathrm{II,I}_\OR(y_k^\o,x_{j,\R}^\pm)\prod_{j=1}^{L_\O} S^\mathrm{II,I}_\OO (y_k^\o,x_{j})~, \qquad k=1,\dots,N_\o~, \\
% 1&=\prod_{j=1}^{L_\L} S^\mathrm{II,I}_\OL (y_k^\op,x_{j,\L}^\pm) \prod_{j=1}^{L_\R} S^\mathrm{II,I}_\OR(y_k^\op,x_{j,\R}^\pm)\prod_{j=1}^{L_\O} S^\mathrm{II,I}_\OO (y_k^\op,x_{j})~, \qquad k=1,\dots,N_\op~, 
% \end{align}
\begin{align}
1&=\prod_{\J \in \{\L,\R,\O\}} \prod_{j=1}^{K_\J} S^\mathrm{I,II}_{\J\L} (x_{\J,j}^\pm,y^-_{\L,k})~, \qquad k=1,\dots,N_\L~, \\
1&=\prod_{\J \in \{\L,\R,\O\}} \prod_{j=1}^{K_\J} S^\mathrm{I,II}_{\J\R} (x_{\J,j}^\pm,y^+_{\R,k})~, \qquad k=1,\dots,N_\R~, \\
1&=\prod_{\J \in \{\L,\R,\O\}} \prod_{j=1}^{K_\J} S^\mathrm{I,II}_{\J\O} (x_{\J,j}^\pm,y^-_{\O,k})~, \qquad k=1,\dots,N_\O~.
\end{align}
% \begin{align}
% S^\mathrm{II,I}_\LL (y, p) &= \frac{D^\LL(y,p)}{A^\LL(y^\L,p)} =-\frac{D^\LL(p,y)}{F^\LL(p,y)}  = e^{\frac{i}{2} p}\frac{y-x^-(p)}{y-x^+(p)}\,,\\
% S^\mathrm{II,I}_\LR (y,p) &= - \frac{E^\LR(y,p)}{C^\LR(y,p)}=  \frac{A^\RL(p,y)}{C^\RL(p,y)} = %e^{-\frac{i}{2} p} \frac{1-y x^+(p)}{1-y x^-(p)}  =
% e^{\frac{i}{2} p} \frac{1-\frac{1}{y\, x^+(p)}}{1-\frac{1}{y\, x^-(p)}}\,,
% \end{align}
% as well as
% \begin{align}
% S^\mathrm{II,I}_\RL (y,p) &= \frac{A^\RL(y,p)}{D^\RL(y,p)} =  -\frac{E^\LR(p,y)}{D^\LR(p,y)} = %e^{\frac{i}{2} p}\frac{1- y x^-(p)}{1-y x^+(p)} =
% e^{-\frac{i}{2} p} \frac{1-\frac{1}{y\, x^-(p)}}{1-\frac{1}{y\, x^+(p)}}\,,\\
% S^\mathrm{II,I}_\RR (y,p) &= - \frac{B^\RR(y,p)}{F^\RR(y,p)}=  \frac{B^\RR(p,y)}{A^\RR(p,y)} = e^{-\frac{i}{2} p} \frac{y- x^+(p)}{y- x^-(p)} \,.
% \end{align}
The eigenvalues are
\begin{align}
\Lambda_\L(x^\pm,\vec{y}) &= +\Delta_\L(x^\pm) \prod_{k=1}^{N_\L} S^\mathrm{I,II}_\LL(x^\pm,y^-_{\L,k}) \prod_{k=1}^{N_\R} S^\mathrm{I,II}_\LR(x^\pm,y_{\R,k}^+)\prod_{k=1}^{N_\O} S^\mathrm{I,II}_\LO(x^\pm,y_{\O,k}^-)\,,\\
\Lambda_\R(x^\pm,\vec{y}) &= -\Delta_\R(x^\pm) \prod_{k=1}^{N_\L} S^\mathrm{I,II}_\RL(x^\pm,y^-_{\L,k}) \prod_{k=1}^{N_\R} S^\mathrm{I,II}_\RR(x^\pm,y_{\R,k}^+)\prod_{k=1}^{N_\O} S^\mathrm{I,II}_\RO(x^\pm,y_{\O,k}^-)\,,\\
\Lambda_\o(x^\pm,\vec{y}) &= +\Delta_\O(x^\pm) \prod_{k=1}^{N_\L} S^\mathrm{I,II}_\OL(x^\pm,y_{\L,k}^-) \prod_{k=1}^{N_\R} S^\mathrm{I,II}_\OR(x^\pm,y_{\R,k}^+)\prod_{k=1}^{N_\O} S^\mathrm{I,II}_\OO(x^\pm,y_{\O,k}^-)\,, \\
\Lambda_\op(x^\pm,\vec{y}) &= -\Delta_\O(x^\pm) \prod_{k=1}^{N_\L} S^\mathrm{I,II}_\OL(x^\pm,y_{\L,k}^-) \prod_{k=1}^{N_\R} S^\mathrm{I,II}_\OR(x^\pm,y_{\R,k}^+)\prod_{k=1}^{N_\O} S^\mathrm{I,II}_\OO(x^\pm,y_{\O,k}^-)~.
\end{align}
%The periodicity condition (here written if only left and right for simplicity) are
%\begin{align}
%-e^{-i p_i^\L L} &= \Lambda_\L(p_i^\L) = \prod_{j=1}^{L_\L}S^\mathrm{I,I}_\LL(p_i^\L,x_{j,\L}^\pm) \prod_{j=1}^{L_\R} S^\mathrm{I,I}_\LR(p_i^\L,x_{j,\R}^\pm)  \prod_{k=1}^{N_\L} S^\mathrm{I,II}_\LL(p_i^\L,y_k^\L) \prod_{k=1}^{N_\R} S^\mathrm{I,II}_\LR(p_i^\L,y_k^\R) \\
%-e^{-i p_i^\R L} &= \Lambda_\R(p_i^\R) = -\prod_{j=1}^{L_\L}S^\mathrm{I,I}_\RL(p_i^\R,x_{j,\L}^\pm) \prod_{j=1}^{L_\R} S^\mathrm{I,I}_\RR(p_i^\R,x_{j,\R}^\pm) \prod_{k=1}^{N_\L} S^\mathrm{I,II}_\RL(p_i^\R,y_k^\L) \prod_{k=1}^{N_\R} S^\mathrm{I,II}_\RR(p_i^\R,y_k^\R)
%\end{align}

\subsection{Simplifying the Bethe equations}
The Bethe equations as well as the eigenvalues only depend on $y_\L^-$, $y_\R^+$ and $y_\O^-$. As we stressed, these are dummy variables. This means that we may reparametrise them as it is more convenient. It is actually convenient~\cite{Borsato:2014hja} to define $y=y_\L^-$ for the ``left'' roots, $y=y_\O^-$ for the ``massless'' roots, as well as $y=1/y_\R^+$ for the ``right'' roots. Observing that 
\begin{align}
\label{eq:auxSfirst}
S^\mathrm{I,II}_\LL(x^\pm,y) &= S^\mathrm{I,II}_\LO(x^\pm,y) = S^\mathrm{I,II}_\LR(x^\pm,1/y) =  \sqrt{\frac{x^-}{x^+}} \frac{y-x^+}{y-x^-} \equiv S^\mathrm{I,II}_{\L \cdot} (x^\pm,y)~\,, \\ 
S^\mathrm{I,II}_\RL(x^\pm,y) &= S^\mathrm{I,II}_\RO(x^\pm,y) = S^\mathrm{I,II}_\RR(x^\pm,1/y) = \sqrt{\frac{x^+}{x^-}} \frac{1-y x^-}{1-y x^+} \equiv S^\mathrm{I,II}_{\R \cdot} (x^\pm,y)\,, \\
S^\mathrm{I,II}_\OL(x^\pm,y) &= S^\mathrm{I,II}_\OO(x^\pm,y) = S^\mathrm{I,II}_\OR(x^\pm,1/y) =  \sqrt{\frac{x^-}{x^+}} \frac{y-x^+}{y-x^-} \equiv S^\mathrm{I,II}_{\O \cdot} (x^\pm,y)\,, 
\label{eq:auxSlast}
\end{align}
all Bethe equations then take the same form
\begin{align}
1&=\prod_{\J\in \{\L,\R,\O\}} \, \prod_{j=1}^{K_\J} S^\mathrm{I,II}_{\J \cdot} (x^\pm_{j},y_k) ~, \qquad k=1,\dots,N~,
\end{align}
with $N=N_\L+N_\R+N_\O$.
The eigenvalues are
\begin{equation}
\Lambda_\mathrm{A}(x^\pm) = (-1)^{\epsilon_\mathrm{A}}\prod_{\mathrm{B}\in \{\L,\R,\O\}} \,  \prod_{j=1}^{K_\mathrm{B}}S^\mathrm{I,I}_{\mathrm{AB}}(x^\pm,x_j^\pm)  \left( 1- \prod_{j=1}^{K_\mathrm{B}}  S^\mathrm{II,I}_{\cdot \mathrm{B}}(x_\mathrm{A},x_j^\pm) \right) \prod_{k=1}^{N} S^\mathrm{I,II}_{\mathrm{A} \cdot}(x^\pm,y_k)\,, 
\end{equation}
where $\mathrm{A} \in \{\text{L},\text{R},\O\}$ and $\epsilon_\L = \epsilon_\o = 0$, $\epsilon_\R=\epsilon_\op=1$. We also used the short-hand $x_\mathrm{A}$, with $x_\L=x^-$, $x_\R=1/x^+$ and $x_\O=x^-=1/x^+$. The main Bethe equations read
\begin{align}
-e^{-i p_i^\mathrm{A} L} &= \Lambda_\mathrm{A}(x^\pm_{i})= (-1)^{\epsilon_\mathrm{A}} \prod_{\mathrm{B}\in \{\L,\R,\O\}} \,  \prod_{j=1}^{K_\mathrm{B}}S^\mathrm{I,I}_{\mathrm{AB}}(x_{i}^\pm,x_{j}^\pm)  \prod_{k=1}^{N} S^\mathrm{I,II}_{\mathrm{A} \cdot}(x_{i}^\pm,y_k)\,,
\end{align}
where again~$\mathrm{A} \in \{\L,\R,\O\}$.
The main S matrices, corresponding to the scattering of highest-weight momentum-carrying modes, are
\begin{align}
S_\LL^\mathrm{I,I}(x_1^\pm,x_2^\pm) &= A^\LL(x_1^\pm,x_2^\pm)~, &\qquad S_\LR^\mathrm{I,I}(x_1^\pm,x_2^\pm)&= C^\LR(x_1^\pm,x_2^\pm)\,,\\
S_\RL^\mathrm{I,I}(x_1^\pm,x_2^\pm) &= D^\RL(x_1^\pm,x_2^\pm)~, &\qquad S_\RR^\mathrm{I,I}(x_1^\pm,x_2^\pm)&= -F^\RR(x_1^\pm,x_2^\pm)\,.
\end{align}
More precisely, these S-matrix elements should be normalised by means of suitable dressing factors. For $AdS_3\times S^3\times T^4$ we need to worry about this normalisation only when considering the ``full'' Bethe equations (which we do in the next section).

\section{Algebraic Bethe Ansatz, full} 
\label{sec:BAfull}

For the full S matrix we consider tensor product of two  representation of $\psu(1|1)^{\oplus2}$ centrally extended. The product of two such representations defines a representation of $\psu(1|1)^{\oplus4}$ c.e.\ only if the two representations of which the tensor product is taken have the same central charges. This construction will give us the Bethe-Yang equations and transfer matrix up to an overall normalisation, which we shall fix following~\cite{Frolov:2021fmj}.

\subsection{Factorised structure}
For the massive representations corresponding to massive fundamental particles in the string theory, we have
\begin{equation}
\varrho_{\mathbb{L}} = \rho_\L \otimes \rho_\L = \{ Y, \Psi^2, \Psi^1, Z \}~, \qquad \varrho_{\mathbb{R}} = \rho_\R \otimes \rho_\R = \{ \tilde{Z}, \tilde{\Psi}^2, \tilde{\Psi}^1, \tilde{Y} \}\,.
\end{equation}
Up to changing the  value of the mass eigenvalue $m$, a similar formula also holds for bound-state representations in the left and right sectors.
There are also two massless representations which we choose to be
\begin{equation}
\varrho_\varoslash = \rho_\o \otimes \rho_\op\,, \qquad \varrho_\varobslash = \rho_\op \otimes \rho_\o\,.
\end{equation}
As a matter of fact, these two representations are isomorphic; they will be distinguished by an $\su(2)$ label $\dot{\alpha}=1,2$; we indicate this symmetry as $\su(2)_\circ$.

In the mirror theory, the discussion is similar with the exception that we are dealing with antisymmetric, rather than symmetric, representations. Practically, this means that the grading of left- and right-representations will be reversed (and, of course, the Zhukovsky variables will need to be analytically continued to the mirror region). For this reason, we will first work in the string theory and then rewrite the equations to obtain the mirror~theory.

The monodromy matrix is a diagonal block matrix, with four $4 \times 4$ blocks,
\begin{equation}
(\MonoB_{\ast})(x^\pm)=(\MonoB_{\star_1 \star_2})(x^\pm) = (\Mono_{\star_1}^{(1)})(x^\pm) \hat{\otimes} (\Mono_{\star_2}^{(2)})(x^\pm)~.
\end{equation}
Here $\ast = \{\mathbb{L},\mathbb{R},\varoslash,\varobslash\}$, while we recall that $\star = \{\L, \R, \o,\op \}$. The identification is such that $\text{LL} \equiv \mathbb{L}$, $\text{RR} \equiv \mathbb{R}$, $\oop \equiv \varoslash$ and $\opo \equiv \varobslash$. All other combinations are forbidden. The superscripts $(1)$ and $(2)$ keep track of the space in the tensor product on which the monodromy matrix acts.~\footnote{These should not be confused with the $\mathfrak{su}(2)$ label distinguishing the two massless representations.} One can then solve the RTT relation for the full problem. Due to the tensor product structure, we find that the monodromy in each copy satisfies the RTT relations of the reduced system, while operators acting on different spaces commute among each other,
\begin{equation}
\label{eq:dum}
\Big[\Mono_{\star_1}^{(1)}(x^\pm),\, \Mono_{\star_2}^{(2)}(y^\pm)\Big]=0\,.
\end{equation}

We define four transfer matrices, one for each sector,
\begin{align}
\TranB_\mathbb{L}(x^\pm) &= +(\MonoB_\mathbb{L})_1{}^1(x^\pm) -  (\MonoB_\mathbb{L})_2{}^2(x^\pm) -  (\MonoB_\mathbb{L})_3{}^3(y^\pm) +  (\MonoB_\mathbb{L})_4{}^4(x^\pm)~, \\
\TranB_\mathbb{R}(x^\pm) &= +(\MonoB_\mathbb{R})_1{}^1(x^\pm) -  (\MonoB_\mathbb{R})_2{}^2(x^\pm) -  (\MonoB_\mathbb{R})_3{}^3(x^\pm) +  (\MonoB_\mathbb{R})_4{}^4(x^\pm)~, \\
\TranB_\varoslash(x^\pm) &= -(\MonoB_\varoslash)_1{}^1(x^\pm) + (\MonoB_\varoslash)_2{}^2(x^\pm) +  (\MonoB_\varoslash)_3{}^3(x^\pm) - (\MonoB_\varoslash)_4{}^4(x^\pm)~, \\
\TranB_\varobslash(x^\pm) &= -(\MonoB_\varobslash)_1{}^1(x^\pm) + (\MonoB_\varobslash)_2{}^2(x^\pm) +  (\MonoB_\varobslash)_3{}^3(x^\pm) - (\MonoB_\varobslash)_4{}^4(x^\pm)~.
\end{align}
There are two ways of thinking about the monodromy and transfer matrix. First we can forget about the tensor product structure and see the monodromy matrix as a big matrix in auxiliary space with four $4 \times 4$ blocks. The tensor product structure then naturally emerges from the commutation (RTT) relations, which take a precise form, traducing the fact that the two copies commute with each other. One then needs to do (a trivial) nesting: if one starts by solving the eigenvalue problem in one copy, the result depends on the eigenvalue of an auxiliary problem which is nothing else than the eigenvalue problem in the other copy. There will thus be two types of auxiliary roots, $y^{(1)}$ and $y^{(2)}$, where the superscript labels the space in the tensor product. Another way of thinking is to make use of the tensor product from the start and define two different creation operators $\MB^{(1)}$ (auxiliary root $y^{(1)}$) and $\MB^{(2)}$ (auxiliary root $y^{(2)}$) acting on each space respectively. We can do this because the nesting is trivial, the two copies commute with each other, and thus solving one problem then the other is equivalent to solve both at the same time, as they do not talk to each other.

\subsection{Vacuum}
We define the vacuum as the tensor product of highest weight states
\begin{equation} 
\ket{\mathbb{0}}  =  \bigotimes_{i=1}^{N_1}\ket{Y(x^\pm_{\L,i})}\bigotimes_{i=1}^{N_{\bar{1}}}\ket{\tilde{Z}(x^\pm_{\R,i})}\bigotimes_{i=1}^{N_{0}^{(1)}}\ket{\chi^1(x^{(1)}_{i})}\bigotimes_{i=1}^{N_{0}^{(2)}}\ket{\chi^2(x^{(2)}_{i})}~.
\end{equation}
% Notice that by definition of the representation we need to impose $K_\mathbb{L}=K_\L^{(1)} =  K_{\L}^{(2)}$, $K_\mathbb{R}=K_\R^{(1)}=K_{\R}^{(2)}$, as well as $K_\varoslash = K^{(1)}_\o = K_{\op}^{(2)}$ and $K_\varobslash = K^{(1)}_\op = K_{\o}^{(2)}$, with the constraint $K_\varoslash + K_\varobslash = K^{(1)}_\o + K^{(1)}_\op = K^{(2)}_{\o}+K^{(2)}_{\op}=K^{(1)}_\O=K^{(2)}_{\O}$. 
This vacuum is an eigenstate of the diagonal entries of the monodromy matrix, with eigenvalues that directly follow from the tensor product structure,
\begin{equation}
    (\MonoB_\ast)_k{}^k(x^\pm) \ket{\mathbb{0}} = \mathbb{\Omega}_{\ast,k}(x^\pm) \ket{\mathbb{0}}\,,\qquad k=1,2,3,4\,,
\end{equation}
with
\begin{equation}
\mathbb{\Omega}_{\ast,2(i-1)+j}(x^\pm) = \mathbb{\Omega}_{\star_1 \star_2,2(i-1)+j}(x^\pm) = \Omega_{\star_1,i}(x^\pm) \Omega_{\star_2,j}(x^\pm)~, \qquad i,j=1,2~.
\end{equation}
 The entries in the lower-triangular part of the monodromy matrix annihilate the vacuum, while the entries in the upper-triangular part can be considered as creation operators. 

For instance, focusing on a vacuum with only $Y$ particles, the operators $(\MonoB_\L)_1{}^2(y) =  \MA_\L^{(1)}(y) \MB_\L^{(2)}(y)$ and $(\MonoB_\L)_3{}^4(y)=\MD_\L^{(1)}(y) \MB_\L^{(2)}(y)$ can be thought of as creating a $\Psi^1$ particle. They are ``proportional'' to each other and in the other perspective mentioned below \eqref{eq:dum} they both correspond to the creation operator $\MB_\L^{(2)}(y^{(2)})$. The operators $(\MonoB_\L)_1{}^3(y) =  \MB_\L^{(1)}(y) \MA_\L^{(2)}(y)$ and $(\MonoB_\L)_2{}^4(y) = \MB_\L^{(1)}(y) \MD_\L^{(2)}(y)$ on the other hand correspond to the operator $\MB_\L^{(1)}(y^{(1)})$ and create a $\Psi^2$ particle. The operator  $(\MonoB_\L)_1{}^4(y)=\MB_\L^{(1)}(y) \MB_\L^{(2)}(y)$ generates a $Z$ particle. In fact this last operator can be though of as creating a double excitation, corresponding to $\MB_\L^{(1)}(y^{(1)}) \MB_\L^{(2)}(y^{(2)})$. The last upper triangular element, $(\MonoB_\L)_2{}^3(y)=\MB_{\L}^{(1)}(y) \MC_\L^{(2)}(y)$ actually annihilates the vacuum. Similar considerations hold for the other sectors.

\subsection{Excited states above the vacuum}
We make the Ansatz
\begin{equation}
\ket{\Phi(\vec{y})} =  \prod_{\star=\L,\R,\o, \op}\, \prod_{\alpha=1}^2 \,\prod_{k=1}^{N_\star^{(\alpha)}}\MB_{\star}^{(\alpha)}(y_{\star,k}^{\pm,(\alpha)}) \ket{\mathbb{0}}\,,
\end{equation}
with $\vec{y}$ a vector gathering all the roots.
This is an eigenstate of the transfer matrices provided one imposes the Bethe equations
\begin{equation}
\label{eq:AuxBetheDelta2}
\Delta_\star(y^{\pm,(\alpha)}_{\star,k}) =0 ~, \qquad k=1,\dots,N_\star^{(\alpha)}~, \qquad \star\in\{\L,\R,\o,\op\}~,
\end{equation}
which upon change of auxiliary variable
\begin{align}
y_{\L,j}^{-,(\alpha)} &= y^{(\alpha)}_j\,, & j&=1,\dots,N_\L^{(\alpha)}\,,  \\
y_{\R,j}^{+,(\alpha)} &= \frac{1}{y^{(\alpha)}_{j+N_\L^{(\alpha)}}}\,, & j&=1,\dots,N_\R^{(\alpha)}\,, \\
y_{\o,j}^{-,(\alpha)} &= y^{(\alpha)}_{j+N_\L^{(\alpha)}+N_\R^{(\alpha)}}\,, & j&=1,\dots,N^{(\alpha)}_\o\,,  \\
y_{\op,j}^{-,(\alpha)} &= y^{(\alpha)}_{j+N_\L^{(\alpha)}+N_\R^{(\alpha)}+N^{(\alpha)}_\o}\,, & j&=1,\dots,N^{(\alpha)}_\op\,, 
\end{align} 
becomes
\begin{align}
1&=\prod_{\star \in \{1,\bar{1},0\}} \prod_{j=1}^{N_{\star}} S^\mathrm{I,II}_{\star\cdot} (x^\pm_{\star,j},y_k^{(\alpha)})~, \qquad k=1,\dots,N_y^{(\alpha)}~, \qquad \alpha=1,2~,
\end{align}
with $S^\mathrm{I,II}_{\star \cdot}(x^\pm,y)$ as in \eqref{eq:auxSfirst}--\eqref{eq:auxSlast} upon identifying $\text{L} \rightarrow 1$, $\text{R} \rightarrow \bar{1}$, $\O \rightarrow 0$ and $N_y^{(\alpha)} = N_\L^{(\alpha)}+N_\R^{(\alpha)}+N^{(\alpha)}_\o+N^{(\alpha)}_\op$.
These are the same as for the factorised S matrix, except that one needs to take into account both sets of auxiliary roots.
The eigenvalues are
\begin{align}\label{eq:eigDelta2}
\mathbb{\Lambda}_\star(x^\pm) &= (-1)^{\tilde{\epsilon}_\star}\left(\Delta_\star(x^\pm)\right)^2 \prod_{\alpha=1}^2\prod_{k=1}^{N_y^{(\alpha)}} S^\mathrm{I,II}_{\star\cdot}(x^\pm,y_k^{(\alpha)})~, \qquad \star\in\{\L,\R,\O\}~,
\end{align}
where $\tilde{\epsilon}_\L=\tilde{\epsilon}_\R=0$ and $\tilde{\epsilon}_\O=1$.
The difference with respect to the result in the factorised case lies in the expectation value on the vacuum, which due to the presence of two copies is now squared, and in presence of two types of auxiliary roots. The dressing factors for the full S matrix are not given by the square of the dressing factors of the factorised S matrix. Hence the $\left(\Delta_\star(x^\pm)\right)^2$ factor shall be modified to obtain the correct eigenvalues.

%with $S^\mathrm{I,I}_{\mathrm{AB}}(x_1,x_2)$ as in \eqref{eq:}.
% The main Bethe equations (periodicity) read
% \begin{align}
% -e^{-i p_i J} =  \mathbb{\Lambda}_\mathrm{A}(x^\pm_{i}) &=  \prod_{\mathrm{J},j}^{L_\mathrm{J}}S^\mathrm{I,I}_\mathrm{A J}(x_{i}^\pm,x_{j}^\pm)^2  \prod_{\alpha=1}^2\prod_{k=1}^{N_y^{(\alpha)}} S^\mathrm{I,II}_{\mathrm{A}\cdot}(x_i^\pm,y_k^{(\alpha)})~, \qquad i =1,\dots, L~, \quad \mathrm{A} \in \{\L,\R,\O\}~.
% \end{align}

\subsection{Normalised Bethe-Yang equations and transfer matrix}
We start by summarising the Bethe equations for fundamental excitations of the string model, and give their explicit form.
It is now important to correctly normalise the various blocks of the S~matrix. For this reason we use the normalisation of~\cite{Frolov:2021fmj}. Following that notation, we write the S matrices as functions of the rapidity $u$ rather than on the Zhukovsky variables $x^\pm$. This is more natural because the dressing factors, unlike the matrix part of the S~matrix, are not any simpler when expressed in terms of Zhukovsky variables; additionally, the rapidity variable is particularly convenient when discussing bound states. We have 
\begin{equation}
\label{eq:boundstateparam}
    x^+_Q+\frac{1}{x^+_Q}-\frac{i\, Q}{h}
    =x^-_Q+\frac{1}{x^-_Q}+\frac{i\, Q}{h}
    =u\,,
\end{equation}
where $Q$ is the bound state number. For left bound-states, the $\mathfrak{u}(1)$ related to the mass is $Q>0$, while for right ones we denote the bound-state number with a bar, $\overline{Q}$, and the $\mathfrak{u}(1)$ charge is~$-\overline{Q}<0$. For massless particles $Q=0$ and we have simply $x+1/x=u$.

The Bethe equations take the following form: 
\paragraph{Left:}
\begin{equation}
1 = e^{i p_i L}\prod_{j \neq i}^{N_1} 
S^{11}_{su}(u_i,u_j)
\prod_{j=1}^{N_{\bar{1}}} \widetilde{S}^{11}_{su}(u_i,u_j)\prod_{\dot{\alpha}=1}^2 \prod_{j=1}^{N_0^{(\dot{\alpha})}} S^{10}(u_i,u_j^{(\dot{\alpha})})\prod_{\alpha=1}^2\prod_{j=1}^{N_y^{(\alpha)}} \overline{S}^{1y}(u_i,y^{(\alpha)}_j).
\end{equation}
\paragraph{Right:}
\begin{equation}
1= e^{i p_i L} \prod_{j=1}^{N_1}  \widetilde{S}^{11}_{sl}(u_i,u_j)
\prod_{j\neq i}^{N_{\bar{1}}} S^{11}_{sl}(u_i,u_j)
\prod_{\dot{\alpha}=1}^2 \prod_{j=1}^{N_0^{(\dot{\alpha})}} 
\bar{S}^{10}(u_i,u_j^{(\dot{\alpha})})
\prod_{\alpha=1}^2\prod_{j=1}^{N_y^{(\alpha)}}
S^{1y}(u_i,y_j^{(\alpha)}).
\end{equation}
\paragraph{Massless:}
\begin{equation}
1 = e^{i p_i L}
\prod_{j=1}^{N_1}
S^{01}(u_i^{(\dot{\alpha})},u_j)
\prod_{j=1}^{N_{\bar{1}}}
\bar{S}^{01}(u_i^{(\dot{\alpha})},u_j)
\prod_{\dot{\beta}=1}^2 \prod_{j=1}^{N_0^{(\dot{\alpha})}}
S^{00}(u_i^{(\dot{\alpha})},u_j^{(\dot{\beta})})
\prod_{\alpha=1}^2\prod_{j=1}^{N_y^{(\alpha)}}
S^{0y}(u_i^{(\dot{\alpha})},y_j^{(\alpha)}).
\end{equation}
\paragraph{Auxiliary:}
\begin{equation}
1 = \prod_{j=1}^{N_1} \overline{S}^{y1}(y_i^{(\alpha)},u_{j}) \prod_{j=1}^{N_{\bar{1}}} S^{y1}(y_i^{(\alpha)},u_{j})
\prod_{\dot{\alpha}=1}^2\prod_{j=1}^{N_0^{(\dot{\alpha})}} S^{y0}(y_i^{(\alpha)},u_j^{(\dot{\alpha})}).
\end{equation}

\paragraph{Diagonalised S~matrices:} We have introduced short-hands for all the diagonal scattering matrices that arise in the Bethe ansatz, and which we will also need for the transfer matrix. Here we followed the naming convention used in~\cite{Frolov:2021bwp}. We have, following the order in which they appeared above:
\begin{align}
S^{11}_{su}(u_i,u_j)&=e^{i p_i}e^{-i p_j} \frac{x^-_i-x^+_j}{x^+_i-x^-_j} \frac{1-\frac{1}{x^-_i x^+_j}}{1-\frac{1}{x^+_i x^-_j}} \left( \sigma^{\bullet\bullet}_{ij}\right)^{-2}\,,
\\
\widetilde{S}^{11}_{su}(u_i,u_j)&=e^{-ip_j} \frac{1-\frac{1}{x^-_ix^-_j}}{1-\frac{1}{x^+_ix^+_j}} \frac{1-\frac{1}{x^-_i x^+_j}}{1-\frac{1}{x^+_i x^-_j}} \left( \tilde{\sigma}^{\bullet\bullet}_{ij}\right)^{-2}\,,
\\
S^{10}(u_i,u_j)&=
e^{\frac{i}{2}p_i}e^{-i p_j} \frac{x_i^--x_j}{1-x_i^+ x_j}  \left( \sigma^{\bullet\circ}_{ij}\right)^{-2}\,,
\\
\overline{S}^{1y}(u,y)&=
e^{-\frac{i}{2} p}\frac{x^+-y}{x^--y}\,.
\end{align}
Then
\begin{align}
\widetilde{S}^{11}_{sl}(u_i,u_j)&=
e^{i p_i} \frac{1-\frac{1}{x^+_ix^+_j}}{1-\frac{1}{x^-_ix^-_j}} \frac{1-\frac{1}{x^-_i x^+_j}}{1-\frac{1}{x^+_i x^-_j}}\left( \tilde{\sigma}^{\bullet\bullet}_{ij}\right)^{-2}\,,
\\
S^{11}_{sl}(u_i,u_j)&=\frac{x_i^+-x_j^-}{x_i^--x_j^+} \frac{1-\frac{1}{x^-_i x^+_j}}{1-\frac{1}{x^+_i x^-_j}} \left( \sigma^{\bullet\bullet}_{ij}\right)^{-2}\,,
\\
\bar{S}^{10}(u_i,u_j)&=
e^{-\frac{i}{2}p_i}e^{-ip_j} \frac{1-x_i^+ x_j}{x_i^--x_j}\left( \sigma^{\bullet\circ}_{ij}\right)^{-2}\\
{S}^{1y}(u,y)&=
e^{\frac{i}{2} p}\frac{1-x^- y }{1- x^+ y}\,.
\end{align}
Next
\begin{align}
S^{01}(u_i,u_j)&=
e^{i p_i}e^{-\frac{i}{2}p_j} \frac{1-x_i x_j^+}{x_i-x_j^-} \left( \sigma^{\circ\bullet}_{ij}\right)^{-2}\,,
\\
\bar{S}^{01}(u_i,u_j)&=
e^{i p_i}e^{\frac{i}{2}p_j} \frac{x_i-x_j^-}{1-x_ix_j^+} \left( \sigma^{\circ\bullet}_{ij}\right)^{-2}\,,
\\
S^{00}(u_i,u_j)&=
\left( \sigma^\OO_{ij}\right)^{-2}\,,
\\
S^{0y}(u,y)&=
e^{-\frac{i}{2} p}\frac{x-y}{1/x-y}\,.
\end{align}
And finally
\begin{align}
\overline{S}^{y1}(y,u)&=\frac{1}{\overline{S}^{1y}(y,u)}
=e^{\frac{i}{2} p}\frac{y-x^-}{y-x^+}\,,
\\
{S}^{y1}(y,u)&=\frac{1}{{S}^{1y}(y,u)} 
=e^{-\frac{i}{2} p} \frac{1-y x^+}{1-y x^-}\,,
\\
S^{y0}(y,u)&=\frac{1}{S^{0y}(y,u)}=
e^{\frac{i}{2} p}\frac{y-1/x}{y-x}\,.
\end{align}
The various dressing factors, indicated by $\sigma$s, are those of~\cite{Frolov:2021fmj}.

Using these expressions, we can write the eigenvalues of the transfer matrix as
\begin{equation} \begin{aligned}
\mathbb{\Lambda}_\L(x^\pm)  &= \prod_{j=1}^{N_1} 
S^{11}_{su}(x^\pm,u_j) \prod_{j=1}^{N_{\bar{1}}} \widetilde{S}^{11}_{su}(x^\pm,u_j) \prod_{\dot{\alpha}=1}^2 \prod_{j=1}^{N_0^{(\dot{\alpha})}} S^{10}(x^\pm,u_j^{(\dot{\alpha})}) \prod_{\alpha=1}^2 \prod_{j=1}^{N_y^{(\alpha)}} \overline{S}^{1y}(x^\pm,y^{(\alpha)}_j) \\
&\qquad \times \left(1-\prod_{j=1}^{N_1}\overline{S}^{y1}(x^-,u_j) \prod_{j=1}^{N_{\bar{1}}}S^{y1}(x^-,u_j)  \prod_{\dot{\alpha}=1}^2 \prod_{j=1}^{N_0^{(\dot{\alpha})}} \overline{S}^{y0}(x^-,u_j^{(\dot{\alpha})}) \right)^2 \,,
\end{aligned}
\end{equation}
\begin{equation} \begin{aligned}
\mathbb{\Lambda}_\R(x^\pm)  &= \prod_{j=1}^{N_1} 
\widetilde{S}^{11}_{sl}(x^\pm,u_j) \prod_{j=1}^{N_{\bar{1}}} S^{11}_{sl}(x^\pm,u_j) \prod_{\dot{\alpha}=1}^2 \prod_{j=1}^{N_0^{(\dot{\alpha})}} \bar{S}^{10}(x^\pm,u_j^{(\dot{\alpha})}) \prod_{\alpha=1}^2 \prod_{j=1}^{N_y^{(\alpha)}} S^{1y}(x^\pm,y^{(\alpha)}_j) \\
&\qquad   \times \left(1-\prod_{j=1}^{N_1}\overline{S}^{y1}(\tfrac{1}{x^+},u_j) \prod_{j=1}^{N_{\bar{1}}}S^{y1}(\tfrac{1}{x^+},u_j)  \prod_{\dot{\alpha}=1}^2 \prod_{j=1}^{N_0^{(\dot{\alpha})}} S^{y0}(\tfrac{1}{x^+},u_j^{(\dot{\alpha})}) \right)^2 \,,
\end{aligned}
\end{equation}
\begin{equation} \begin{aligned}
\mathbb{\Lambda}_\O(x^\pm)  &= - \prod_{j=1}^{N_1} 
S^{01}(x^\pm,u_j)  \prod_{j=1}^{N_{\bar{1}}} \bar{S}^{01}(x^\pm,u_j) \prod_{\dot{\alpha}=1}^2 \prod_{j=1}^{N_0^{(\dot{\alpha})}} S^{00}(x^\pm,u_j^{(\dot{\alpha})}) \prod_{\alpha=1}^2 \prod_{j=1}^{N_y^{(\alpha)}} S^{0y}(x^\pm,y^{(\alpha)}_j) \\
&\qquad 
  \times \left(1-\prod_{j=1}^{N_1}\overline{S}^{y1}(x^-,u_j) \prod_{j=1}^{N_{\bar{1}}}S^{y1}(x^-,u_j)  \prod_{\dot{\alpha}=1}^2 \prod_{j=1}^{N_0^{(\dot{\alpha})}} S^{y0}(x^-,u_j^{(\dot{\alpha})}) \right)^2 \,,
\end{aligned}
\end{equation}
where in the last equation we may also write $x=x^+=1/x^-$.

\subsection{Transfer matrix for the string-theory bound states}
We can now extend our discussion to the bound states of the string theory. For left particles, a generic $Q$-particle bound states is given by the condition
\begin{equation}
\label{eq:Qboundstate}
    x^+_{j}=x^-_{j+1}\,,\qquad
    j=1,\dots, Q-1\,.
\end{equation}
This condition can be solved in terms of the~$u$ parametrisation, so that the bound state depends on~\eqref{eq:boundstateparam}. The bound state S~matrices can be constructed by fusion. For instance, the auxiliary S~matrix for a $Q$-particle bound state is
\begin{equation}
    S^{Qy}(u,y)=\prod_{j=1}^Q S(u_j,y)=e^{-\tfrac{i}{2}p_Q}\frac{x^+_Q-y}{x^-_Q-y}\,,
\end{equation}
where the product is taken over the configuration~\eqref{eq:Qboundstate}.
For right particles, however, the description of the bound state is a little more awkward. The reason is that in the diagonalisation of the S~matrix we have picked as reference vacuum state the excitation $|\widetilde{Z}\rangle$. However, in string theory we are dealing with symmetric bound states, which are constructed out of several copies of~$|\widetilde{Y}\rangle$. For this reason, we need to consider a configuration involving a string of~$\overline{Q}$ ``right'' excitations satisfying~\eqref{eq:Qboundstate} (with $j=1,\dots, \overline{Q}$), as well as $2(\overline{Q}-1)$ auxiliary particles, satisfying $y^{(\alpha)}_j=1/x^+_j$,  with $j=1,\dots, \overline{Q}-1$ and $\alpha=1,2$.
Finally, for massless particles and auxiliary~$y$ particles there are no bound-states, see~\cite{Frolov:2021fmj}, where the fusion properties of the dressing factors in the string region is also discussed.
The explicit form of the fused S~matrices can be found in~\cite{Frolov:2021bwp}; for convenience, we are following the same naming convention as it is used there.

Taking these considerations into account, it is not difficult to write the bound-state transfer matrices as (the index $a$ runs on all left or right bound states, of bound-state number $Q_a$ and $\overline{Q}_a$, respectively)
\begin{equation} \begin{aligned}
\mathbb{\Lambda}_Q(x^\pm)  &= \prod_{a\in\text{L}}
S^{QQ_a}_{su}(x^\pm,u_a)\prod_{a\in\text{R}} \widetilde{S}^{Q\overline{Q}_a}_{su}(x^\pm,u_a) \prod_{\dot{\alpha}=1}^2 \prod_{j=1}^{N_0^{(\dot{\alpha})}} S^{Q0}(x^\pm,u_j^{(\dot{\alpha})}) \prod_{\alpha=1}^2 \prod_{j=1}^{N_y^{(\alpha)}} \overline{S}^{Qy}(x^\pm,y^{(\alpha)}_j) \\
&\qquad  \left(1-\prod_{a\in\text{L}} \overline{S}^{yQ_a}(x^-,u_a) \prod_{a\in\text{R}} S^{y\overline{Q}_a}(x^-,u_a)   \prod_{\dot{\alpha}=1}^2 \prod_{j=1}^{N_0^{(\dot{\alpha})}} S^{y0}(x^-,u_j^{(\dot{\alpha})}) \right)^2 \,,
\end{aligned}
\end{equation}
\begin{equation} \begin{aligned}
\mathbb{\Lambda}_{\overline{Q}}(x^\pm)  &= \prod_{a\in\text{L}} 
\widetilde{S}^{\overline{Q}Q_a}_{sl}(x^\pm,u_a) \prod_{a\in\text{R}} S^{\overline{QQ}_a}_{sl}(x^\pm,u_a) \prod_{\dot{\alpha}=1}^2 \prod_{j=1}^{N_0^{(\dot{\alpha})}} \bar{S}^{\overline{Q}0}(x^\pm,u_j^{(\dot{\alpha})}) \prod_{\alpha=1}^2 \prod_{j=1}^{N_y^{(\alpha)}} S^{\overline{Q}y}(x^\pm,y^{(\alpha)}_j)  \\
&\qquad  \left(1-\prod_{a\in\text{L}} \overline{S}^{yQ_a}(\tfrac{1}{x^+},u_a) \prod_{a\in\text{R}} S^{y\overline{Q}_a}(\tfrac{1}{x^+},u_a)   \prod_{\dot{\alpha}=1}^2 \prod_{j=1}^{N_0^{(\dot{\alpha})}} S^{y0}(\tfrac{1}{x^+},u_j^{(\dot{\alpha})}) \right)^2 \,,
\end{aligned}
\end{equation}
\begin{equation} \begin{aligned}
\mathbb{\Lambda}_\O(x)  &= - \prod_{a\in\text{L}} 
S^{0Q_a}(x,u_a)  \prod_{a\in\text{R}} \bar{S}^{0\overline{Q}_a}(x,u_a) \prod_{\dot{\alpha}=1}^2 \prod_{j=1}^{N_0^{(\dot{\alpha})}} S^{00}(x,u_j^{(\dot{\alpha})}) \prod_{\alpha=1}^2 \prod_{j=1}^{N_y^{(\alpha)}} S^{0y}(x,y^{(\alpha)}_j) \\
&\qquad \left(1-\prod_{a\in\text{L}} \overline{S}^{yQ_a}(\tfrac{1}{x},u_a) \prod_{a\in\text{R}} S^{y\overline{Q}_a}(\tfrac{1}{x},u_a)   \prod_{\dot{\alpha}=1}^2 \prod_{j=1}^{N_0^{(\dot{\alpha})}} S^{y0}(\tfrac{1}{x},u_j^{(\dot{\alpha})}) \right)^2  \,.
\end{aligned}
\end{equation}
The Bethe equations are
\paragraph{Left:}
\begin{equation}
1 = e^{i p_b L}\prod_{a \in \L, a \neq b}
S^{QQ_a}_{su}(u_b,u_a)
\prod_{a \in \R} \widetilde{S}^{Q \overline{Q}_a}_{su}(u_b,u_a)\prod_{\dot{\alpha}=1}^2 \prod_{j=1}^{N_0^{(\dot{\alpha})}} S^{Q0}(u_b,u_j^{(\dot{\alpha})})\prod_{\alpha=1}^2\prod_{j=1}^{N_y^{(\alpha)}} \overline{S}^{Qy}(u_b,y^{(\alpha)}_j).
\end{equation}
\paragraph{Right:}
\begin{equation}
1= e^{i p_b L} \prod_{a \in \L}  \widetilde{S}^{\overline{Q} Q_a}_{sl}(u_b,u_a)
\prod_{a \in \R, a \neq b} S^{\overline{Q} \overline{Q}_a}_{sl}(u_b,u_a)
\prod_{\dot{\alpha}=1}^2 \prod_{j=1}^{N_0^{(\dot{\alpha})}} 
\bar{S}^{\overline{Q}0}(u_b,u_j^{(\dot{\alpha})})
\prod_{\alpha=1}^2\prod_{j=1}^{N_y^{(\alpha)}}
S^{\overline{Q}y}(u_b,y_j^{(\alpha)}).
\end{equation}
\paragraph{Massless:}
\begin{equation}
1 = e^{i p_i L}
\prod_{a \in \L}
S^{0 Q_a}(u_i^{\dot{\alpha}},u_a)
\prod_{a \in \R}
\bar{S}^{0\overline{Q}_a}(u_i,u_a)
\prod_{\dot{\beta}=1}^2 \prod_{j=1}^{N_0^{(\dot{\beta})}}
S^{00}(u_i^{\dot{\alpha}},u_j^{(\dot{\beta})})
\prod_{\alpha=1}^2\prod_{k=1}^{N_y^{(\alpha)}}
S^{0y}(u_i^{\dot{\alpha}},y_j^{(\alpha)}).
\end{equation}
\paragraph{Auxiliary:}
\begin{equation}
1 = \prod_{a \in \L} \overline{S}^{yQ}(y_i^{(\alpha)},u_a) \prod_{a \in \R} S^{y\overline{Q}}(y_i^{(\alpha)},u_a)
\prod_{\dot{\alpha}=1}^2\prod_{j=1}^{N_0^{(\dot{\alpha})}} S^{y0}(y_i^{(\alpha)},u_j^{(\dot{\alpha})}).
\end{equation}

\subsection{Transfer matrix for the mirror theory}
We have argued that up to swapping the grading of left and right excitations, and up to analytic continuation of the Zhukovsky variables, the matrix part of the S~matrix is the same in the string and mirror theory. More precisely, in the string theory the left particles were in the $\mathfrak{su}(2)$ grading, which forced the right-particles to be in the $\mathfrak{sl}(2)$ grading. As a result, in the string theory  bound-states of left particles would have a simple expression, contrary to those of right-particles. In the mirror theory, left particles will be in the $\mathfrak{sl}(2)$ grading, and right particles will be in the $\mathfrak{su}(2)$ grading. This means that mirror bound states will again have a simple description in the case of left particles (and again a more awkward description in the case of right particles).
As a result, the mirror Bethe-Yang equations coincide perfectly with those used in~\cite{Frolov:2021bwp} and they are
\begin{equation}
1 = e^{i \tilde{p}_i R}\prod_{j \neq i}^{N_1} 
S^{11}_{sl}(u_i,u_j)
\prod_{j=1}^{N_{\bar{1}}} \widetilde{S}^{11}_{sl}(u_i,u_j)\prod_{\dot{\alpha}=1}^2 \prod_{j=1}^{N_0^{(\dot{\alpha})}} S^{10}(u_i,u_j^{(\dot{\alpha})})\prod_{\alpha=1}^2\prod_{j=1}^{N_y^{(\alpha)}} S^{1y}(u_i,y^{(\alpha)}_j).
\end{equation}
\begin{equation}
1= e^{i \tilde{p}_i R} \prod_{j=1}^{N_1}  \widetilde{S}^{11}_{su}(u_i,u_j)
\prod_{j\neq i}^{N_{\bar{1}}} S^{11}_{su}(u_i,u_j)
\prod_{\dot{\alpha}=1}^2 \prod_{j=1}^{N_0^{(\dot{\alpha})}} 
\bar{S}^{10}(u_i,u_j^{(\dot{\alpha})})
\prod_{\alpha=1}^2\prod_{j=1}^{N_y^{(\alpha)}}
\overline{S}^{1y}(u_i,y_j^{(\alpha)}).
\end{equation}
\begin{equation}
\label{eq:masslessmirror}
-1 = e^{i \tilde{p}_i R}
\prod_{j=1}^{N_1}
S^{01}(u_i^{\dot{\alpha}},u_j)
\prod_{j=1}^{N_{\bar{1}}}
\bar{S}^{01}(u_i^{\dot{\alpha}},u_j)
\prod_{\dot{\beta}=1}^2 \prod_{j=1}^{N_0^{(\dot{\beta})}}
S^{00}(u_i^{\dot{\alpha}},u_j^{(\dot{\beta})})
\prod_{\alpha=1}^2\prod_{k=1}^{N_y^{(\alpha)}}
S^{0y}(u_i^{\dot{\alpha}},y_j^{(\alpha)}).
\end{equation}
\begin{equation}
\label{eq:auxmirror}
-1 = \prod_{j=1}^{N_1} S^{y1}(y_i^{(\alpha)},u_{j}) \prod_{j=1}^{N_{\bar{1}}} \overline{S}^{y1}(y_i^{(\alpha)},u_{j})
\prod_{\dot{\alpha}=1}^2\prod_{j=1}^{N_0^{(\dot{\alpha})}} S^{y0}(y_i^{(\alpha)},u_j^{(\dot{\alpha})}).
\end{equation}
We have a few differences with the string Bethe-Yang equations. First of all, the mirror length is~$R$ (rather than~$L$), and the volume phase-factor is given in terms of the mirror momentum~$\tilde{p}$. The S-matrix elements are continued to the mirror region, see~\cite{Frolov:2021fmj,Frolov:2021bwp}.
Additionally, we have an additional minus sign in~\eqref{eq:masslessmirror} and one in~\eqref{eq:auxmirror} due to the fact that Fermions are antiperiodic.

\paragraph{Mirror transfer matrix, fundamental particles.}
The mirror transfer matrix follows straightforwardly from the string one. We report it here for completeness.
\begin{equation} \begin{aligned}
\mathbb{\Lambda}_\L(x^\pm)  &= \prod_{j=1}^{N_1} 
S^{11}_{sl}(x^\pm,u_j) \prod_{j=1}^{N_{\bar{1}}} \widetilde{S}^{11}_{sl}(x^\pm,u_j) \prod_{\dot{\alpha}=1}^2 \prod_{j=1}^{N_0^{(\dot{\alpha})}} S^{10}(x^\pm,u_j^{(\dot{\alpha})}) \prod_{\alpha=1}^2 \prod_{j=1}^{N_y^{(\alpha)}} S^{1y}(x^\pm,y^{(\alpha)}_j) \\
&\qquad \times \left(1-\prod_{j=1}^{N_1}S^{y1}(x^-,u_j) \prod_{j=1}^{N_{\bar{1}}}\overline{S}^{y1}(x^-,u_j)  \prod_{\dot{\alpha}=1}^2 \prod_{j=1}^{N_0^{(\dot{\alpha})}} S^{y0}(x^-,u_j^{(\dot{\alpha})}) \right)^2 \,,
\end{aligned}
\end{equation}
\begin{equation} \begin{aligned}
\mathbb{\Lambda}_\R(x^\pm)  &= \prod_{j=1}^{N_1} 
\widetilde{S}^{11}_{su}(x^\pm,u_j) \prod_{j=1}^{N_{\bar{1}}} S^{11}_{su}(x^\pm,u_j) \prod_{\dot{\alpha}=1}^2 \prod_{j=1}^{N_0^{(\dot{\alpha})}} \bar{S}^{10}(x^\pm,u_j^{(\dot{\alpha})}) \prod_{\alpha=1}^2 \prod_{j=1}^{N_y^{(\alpha)}} \overline{S}^{1y}(x^\pm,y^{(\alpha)}_j) \\
&\qquad   \times \left(1-\prod_{j=1}^{N_1}S^{y1}(\tfrac{1}{x^+},u_j) \prod_{j=1}^{N_{\bar{1}}}\overline{S}^{y1}(\tfrac{1}{x^+},u_j)  \prod_{\dot{\alpha}=1}^2 \prod_{j=1}^{N_0^{(\dot{\alpha})}} S^{y0}(\tfrac{1}{x^+},u_j^{(\dot{\alpha})}) \right)^2 \,,
\end{aligned}
\end{equation}
\begin{equation} \begin{aligned}
\mathbb{\Lambda}_\O(x^\pm)  &= - \prod_{j=1}^{N_1} 
S^{01}(x^\pm,u_j)  \prod_{j=1}^{N_{\bar{1}}} \bar{S}^{01}(x^\pm,u_j) \prod_{\dot{\alpha}=1}^2 \prod_{j=1}^{N_0^{(\dot{\alpha})}} S^{00}(x^\pm,u_j^{(\dot{\alpha})}) \prod_{\alpha=1}^2 \prod_{j=1}^{N_y^{(\alpha)}} S^{0y}(x^\pm,y^{(\alpha)}_j) \\
&\qquad 
  \times \left(1-\prod_{j=1}^{N_1}S^{y1}(x^-,u_j) \prod_{j=1}^{N_{\bar{1}}}\overline{S}^{y1}(x^-,u_j)  \prod_{\dot{\alpha}=1}^2 \prod_{j=1}^{N_0^{(\dot{\alpha})}} S^{y0}(x^-,u_j^{(\dot{\alpha})}) \right)^2 \,.
\end{aligned}
\end{equation}

\paragraph{Mirror transfer matrix, bound states.}
Coming now to the bound states, their mirror Bethe-Yang equations were derived in~\cite{Frolov:2021bwp} in some detail, and we refer the reader to that article for their explicit form (remark that our notation coincides with theirs). As for the bound-state transfer matrix, we have
\begin{equation} \begin{aligned}
\mathbb{\Lambda}_Q(x^\pm)  &= \prod_{a\in\text{L}}
S^{QQ_a}_{sl}(x^\pm,u_a)\prod_{a\in\text{R}} \widetilde{S}^{Q\overline{Q}_a}_{sl}(x^\pm,u_a) \prod_{\dot{\alpha}=1}^2 \prod_{j=1}^{N_0^{(\dot{\alpha})}} S^{Q0}(x^\pm,u_j^{(\dot{\alpha})}) \prod_{\alpha=1}^2 \prod_{j=1}^{N_y^{(\alpha)}} S^{Qy}(x^\pm,y^{(\alpha)}_j) \\
&\qquad  \left(1-\prod_{a\in\text{L}} S^{yQ_a}(x^-,u_a) \prod_{a\in\text{R}} \overline{S}^{y\overline{Q}_a}(x^-,u_a)   \prod_{\dot{\alpha}=1}^2 \prod_{j=1}^{N_0^{(\dot{\alpha})}} S^{y0}(x^-,u_j^{(\dot{\alpha})}) \right)^2 \,,
\end{aligned}
\end{equation}
\begin{equation} \begin{aligned}
\mathbb{\Lambda}_{\overline{Q}}(x^\pm)  &= \prod_{a\in\text{L}} 
\widetilde{S}^{\overline{Q}Q_a}_{su}(x^\pm,u_a) \prod_{a\in\text{R}} S^{\overline{QQ}_a}_{su}(x^\pm,u_a) \prod_{\dot{\alpha}=1}^2 \prod_{j=1}^{N_0^{(\dot{\alpha})}} \bar{S}^{\overline{Q}0}(x^\pm,u_j^{(\dot{\alpha})}) \prod_{\alpha=1}^2 \prod_{j=1}^{N_y^{(\alpha)}} \overline{S}^{\overline{Q}y}(x^\pm,y^{(\alpha)}_j)  \\
&\qquad  \left(1-\prod_{a\in\text{L}} S^{yQ_a}(\tfrac{1}{x^+},u_a) \prod_{a\in\text{R}} \overline{S}^{y\overline{Q}_a}(\tfrac{1}{x^+},u_a)   \prod_{\dot{\alpha}=1}^2 \prod_{j=1}^{N_0^{(\dot{\alpha})}} S^{y0}(\tfrac{1}{x^+},u_j^{(\dot{\alpha})}) \right)^2 \,,
\end{aligned}
\end{equation}
\begin{equation} \begin{aligned}
\mathbb{\Lambda}_\O(x)  &= - \prod_{a\in\text{L}} 
S^{0Q_a}(x,u_a)  \prod_{a\in\text{R}} \bar{S}^{0\overline{Q}_a}(x,u_a) \prod_{\dot{\alpha}=1}^2 \prod_{j=1}^{N_0^{(\dot{\alpha})}} S^{00}(x,u_j^{(\dot{\alpha})}) \prod_{\alpha=1}^2 \prod_{j=1}^{N_y^{(\alpha)}} S^{0y}(x,y^{(\alpha)}_j) \\
&\qquad \left(1-\prod_{a\in\text{L}} S^{yQ_a}(\tfrac{1}{x},u_a) \prod_{a\in\text{R}} \overline{S}^{y\overline{Q}_a}(\tfrac{1}{x},u_a)   \prod_{\dot{\alpha}=1}^2 \prod_{j=1}^{N_0^{(\dot{\alpha})}} S^{y0}(\tfrac{1}{x},u_j^{(\dot{\alpha})}) \right)^2  \,.
\end{aligned}
\end{equation}

\section{Twisted transfer matrix}
\label{sec:twists}
We are now interested in discussing the Bethe-Yang equations and transfer matrix when the model is twisted. By this we mean that we want to consider a Bosonic twist $G$ which commutes with the action of the (full) S~matrix,
\begin{equation}
    [S(p_1,p_2),G\otimes G]=0\,.
\end{equation}
One possibility is that $G$ is a combination of $PSU(1,1|2)^{\otimes2}$ bosonic generators. If that is the case, $G$ must actually be a combination of the Cartan elements of $PSU(1,1|2)^{\otimes2}$, as all other Bosonic generators do not commute with the S~matrix. Another possibility is that the twist is in $SU(2)_\bullet\otimes SU(2)_\circ$, which geometrically represent rotations on $T^4$ and corresponds to the labels $\alpha=1,2$ (for auxiliary roots~$y^{(\alpha)}$) and $\dot{\alpha}=1,2$ (for massless roots~$x^{(\dot{\alpha})}$), respectively~\cite{Borsato:2014hja}. In this case, it is sufficient to consider the Cartan elements of each~$SU(2)$, see~\cite{deLeeuw:2012hp}. In conclusion, without loss of generality we are dealing with a \textit{diagonal} twist.

\subsection{Parametrising the diagonal twists}
\label{sec:twists:diagonal}
We now want to parametrise how the twist acts on fundamental particles in the string model. Again, a new element with respect to \textit{e.g.}~\cite{deLeeuw:2012hp} is that here we are dealing with several reducible representations.

As a warm up, let us discuss the baby example of the $\rho_\L^\B$ and $\rho_\R^\F$ representations, which are two-dimensional. We will work in the basis
\begin{equation}
    \left(\ket{\phi^\B_\L},\ket{\varphi^\F_\L};\ 
    \ket{\phi^\F_\R},\ket{\varphi^\B_\R}\right)\,.
\end{equation}
The most general twist that we may consider depends on four parameters, and we may write it for instance as
\begin{equation}
    G=\left(\begin{array}{cc|cc}
    e^{+\tfrac{i}{2} \mu_\L}e^{+\tfrac{i}{2} \nu_\L} & 0 & 0 & 0\\
    0 & e^{+\tfrac{i}{2} \mu_\L}e^{-\tfrac{i}{2} \nu_\L} & 0 & 0\\
    \hline
    0 & 0 & e^{+\tfrac{i}{2} \mu_\R}e^{+\tfrac{i}{2} \nu_\R} & 0\\
    0 & 0 & 0 & e^{+\tfrac{i}{2} \mu_\R}e^{-\tfrac{i}{2} \nu_\R}
    \end{array}\right)\,.
\end{equation}
It is easy to see that not all these parameters may be unrelated. In particular, demanding commutation with the left-right scattering matrix we find
\begin{equation}
    \nu_\L=\nu_\R\,.
\end{equation}
In other words, the parameter $\nu=\nu_\L=\nu_\R$ measures whether we are in the highest- or lowest-weight states, independently from the representation. This observation is perfectly consistent with the fact that, even if we have several reducible representations, we only have one set of $y$-roots per supercharge (see appendix~\ref{app:Null}).

Let us now consider a four-dimensional representation of $\psu(1|1)^{\oplus4}$ centrally extended, for instance the left one, whose basis is
\begin{equation}
    \left(\ket{Y},\ket{\Psi^{\alpha}},\ket{Z}\right)\,.
\end{equation}
The charge under $\mu_\L$ and $\nu$ follows from the tensor-product structure. Additionally, we may introduce a new charge~$\rho$ which corresponds to the Cartan of $SU(2)_\bullet$ and distinguishes the two states $\ket{\Psi^{\alpha}}$. All in all we may write
\begin{equation}
    G=e^{i \mu_\L}
    \left(\begin{array}{cccc}
    e^{+i \nu} & 0 & 0 & 0\\
    0 & e^{+\tfrac{i}{2} \rho} & 0 & 0\\
    0 & 0 & e^{-\tfrac{i}{2} \rho} & 0\\
    0 & 0 & 0 & e^{-i \nu}
    \end{array}\right)\,.
\end{equation}
For the right representation, in the basis
\begin{equation}
    \left(\ket{\tilde{Z}},\ket{\tilde{\Psi}^{\alpha}},\ket{\tilde{Y}}\right)\,,
\end{equation}
we write
\begin{equation}
    G=e^{i \mu_\R}
    \left(\begin{array}{cccc}
    e^{+i \nu} & 0 & 0 & 0\\
    0 & e^{+\tfrac{i}{2} \rho} & 0 & 0\\
    0 & 0 & e^{-\tfrac{i}{2} \rho} & 0\\
    0 & 0 & 0 & e^{-i \nu}
    \end{array}\right)\,,
\end{equation}
and for the two massless representation in the basis
\begin{equation}
    \left(\ket{\chi^{\dot{\alpha}}},\ket{T^{\dot{\alpha}\alpha}},\ket{\tilde{\chi}^{\dot{\alpha}}}\right)\,,
\end{equation}
we write
\begin{equation}
    G=e^{i \mu_\circ}e^{\pm\tfrac{i}{2}\dot{\rho}}
    \left(\begin{array}{cccc}
    e^{+i \nu} & 0 & 0 & 0\\
    0 & e^{+\tfrac{i}{2} \rho} & 0 & 0\\
    0 & 0 & e^{-\tfrac{i}{2} \rho} & 0\\
    0 & 0 & 0 & e^{-i \nu}
    \end{array}\right)\,,
\end{equation}
where the chemical potential~$\dot{\rho}$ corresponds to $SU(2)_\circ$ and distinguishes the cases~$\dot{\alpha}=1,2$.

\subsection{Identifying the physical charges}
The twisted transfer matrix and Bethe-Yang equations may be used to study interesting models such as the TsT-transformed background or orbifold models, see \textit{e.g.}~\cite{vanTongeren:2013gva} for a review. To this end, it is important to relate the chemical potentials introduced above to the actual symmetries of the model.

Two of these chemical potentials have an obvious interpretation. The parameters $\rho$ and $\dot{\rho}$ are clearly related to the (Cartan element of) the $SU(2)_\circ$ and $SU(2)_\bullet$ symmetries, which come from the $T^4$ directions and give charges to all Fermions in the model (as well as to the $T^4$ bosons, which are bispinors under $SU(2)_\circ\otimes SU(2)_\bullet$). The twist in the transfer matrix comes from
\begin{equation}
    e^{i\, \dot{\rho}\,\gen{J}_\circ}\,
    e^{i\, \rho\,\gen{J}_\bullet}\,.
\end{equation}

It is a little less obvious to identify the chemical potentials related to the twists on $AdS_3\times S^3$. These should be associated to the Cartan elements of $SO(2,2)\otimes SO(4)\subset PSU(1,1|2)^{\otimes 2}$. These Cartan elements are
\begin{equation}
    \gen{L}_0\,,\qquad\widetilde{\gen{L}}_0\,,\qquad
    \gen{J}^3\,,\qquad\widetilde{\gen{J}}^3\,,
\end{equation}
where the first two are related to AdS and the latter two to the sphere. However, not all of these combinations commute with the lightcone symmetry algebra. In particular, the total light-cone momentum
\begin{equation}
\label{eq:Pplus}
  \gen{P}_+ = \gen{L}_0+\widetilde{\gen{L}}_0+\gen{J}^3+\widetilde{\gen{J}}^3\,,
\end{equation}
gives the worldsheet size in lightcone gauge and has non-trivial commutation relation with the central extension~$\gen{C}$, see~\cite{Arutyunov:2006ak} for a detailed discussion of this point. In the construction of the S~matrix, one typically takes $\gen{P}_+=\infty$, and decouples this generator. Twists involving~$\gen{P}_+$ have recently been explored in~\cite{vanTongeren:2021jhh} and they result in a non-trivial interplay of the target-space and light-cone variables similarly to $T\bar{T}$ deformations on the worldsheet~\cite{Sfondrini:2019smd}. This discussion is beyond the scope of our paper, and we will restrict ourselves to ``traditional'' twists that commute with the lightcone symmetry algebra.

We can parametrise the space of the generators orthogonal to~\eqref{eq:Pplus} in terms of
\begin{equation}
    \gen{H}=\gen{L}_0-\gen{J}^3\,,\qquad
    \widetilde{\gen{H}}=\widetilde{\gen{L}}_0-\widetilde{\gen{J}}^3\,,\qquad
    \gen{s}=
    \gen{L}_0-\widetilde{\gen{L}}_0+\gen{J}^3-\tilde{\gen{J}}^3\,.
\end{equation}
The first two are the left- and right-Hamiltonian. They are not quantised, but receive quantum corrections (their difference, however, is quantised). The latter is a total spin on $AdS_3\times S^3$ and it is quantised.

Let us start by discussing~$\nu$, which was related to the action of the lowering operators $\gen{Q}^{\alpha}$ and~$\widetilde{\gen{S}}^{\alpha}$. Because the lowering operators commute with~$\gen{H}$ and~$\widetilde{\gen{H}}$, it follows that $\nu$ can only be related to~$\gen{s}$. Looking more closely at our definition we find that, for any of the states above, the chemical potentials arise precisely by acting with
\begin{equation}
    e^{i\, \nu\, \gen{s}}\,.
\end{equation}

Finally, we have the parameters~$\mu_\star$. We note that we seemingly have three conserved parameters for two-conserved charges. Essentially, looking at the symmetries of the S~matrix we would imagine that the number of left, right and massless excitation is separately conserved. Instead, looking at the charges we have
\begin{equation}
    e^{i\, \mu\, \gen{H}}\,
    e^{i\, \tilde{\mu}\, \widetilde{\gen{H}}}\,,
\end{equation}
which can be mapped back to the~$\mu_\star$ parameters as it follows
\begin{equation}
    \mu_\L=\frac{\mu}{2}(2L_\L+\delta H)+\frac{\tilde{\mu}}{2}\delta H,\qquad
    \mu_\R=\frac{\mu}{2}\delta H+\frac{\tilde{\mu}}{2}(2L_\R+\delta H),\qquad
    \mu_\circ=\frac{\mu+\tilde{\mu}}{2}\delta H,
\end{equation}
where $\delta H$ is the ``anomalous dimensions'' due to the presence of RR background fluxes (meaning, it vanishes at $h=0$)
\begin{equation}
    \delta H = \sum_j \sqrt{m^2_j+4h^2\sin^2(\tfrac{1}{2}p_j)}-|m_j|\,.
\end{equation}
All in all, the twist is generated by the operator
\begin{equation}
    \gen{G} = e^{i\, \mu\, \gen{H}}\,
    e^{i\, \tilde{\mu}\, \widetilde{\gen{H}}}\,e^{i\, \nu\,\gen{s}}\,
    e^{i\, \dot{\rho}\,\gen{J}_\circ}\,
    e^{i\, \rho\,\gen{J}_\bullet}\,.
\end{equation}

We summarise the charges of the string and mirror bound states under the various Cartan generators in tables~\ref{tab:stringcharges} and~\ref{tab:mirrorcharges}, respectively.

\begin{table}[t]
    \centering
    \begin{tabular}{c|c|c|c|c|c}
    & $\gen{L}_0$ &  $\widetilde{\gen{L}}_0$ & $\gen{J}^3$ & $\widetilde{\gen{J}}^3 $& $\gen{s}$ \\ \hline
    $Y$    &  $\delta H/2$  & $\delta H/2$ & $-Q_s$ & 0 & $-Q_s$\\
    $\Psi^{\alpha}$ & $\delta H/2+ 1/2$ & $\delta H/2$ & $-Q_s+1/2$ & 0 & $-Q_s+1$ \\
    $Z$ & $\delta H/2+1$ & $\delta H/2$ & $-Q_s+1$ & 0 & $-Q_s+2$ \\ \hline
    $\tilde{Z}$    &  $\delta H/2$  & $\delta H/2+1$ & 0 & $-\overline{Q}_s+1$ & $\overline{Q}_s-2$\\
    $\tilde{\Psi}^{\alpha}$ & $\delta H/2$ & $\delta H/2+1/2$ & 0 & $-\overline{Q}_s+1/2$ & $\overline{Q}_s-1$ \\
    $\tilde{Y}$ & $\delta H/2$ & $\delta H/2$ & 0 & $-\overline{Q}_s$ & $\overline{Q}_s$ \\ \hline
    \end{tabular}
    \caption{Charges of the massive particles and their string theory bound states under the Cartan elements. $Q_s$ denotes the mass of the left bound state, with $Q_s=1$ for fundamental particles. Similarly, $\overline{Q}_s$ denotes the mass of the right bound state, with $\overline{Q}_s=1$ for fundamental particles.}
    \label{tab:stringcharges}
\end{table}
\begin{table}[t]
    \centering
    \begin{tabular}{c|c|c|c|c|c}
         & $\gen{L}_0$ &  $\widetilde{\gen{L}}_0$ & $\gen{J}^3$ & $\widetilde{\gen{J}}^3 $& $\gen{s}$ \\ \hline
    $Y$    &  $\delta H/2+Q_m-1$  & $\delta H/2$ & $-1$ & 0 & $Q_m-2$\\
    $\Psi^{\alpha}$ & $\delta H/2+ Q_m-1/2$ & $\delta H/2$ & $-1/2$ & 0 & $Q_m-1$ \\
    $Z$ & $\delta H/2+Q_m$ & $\delta H/2$ & $0$ & 0 & $Q_m$ \\ \hline
    $\tilde{Z}$    &  $\delta H/2$  & $\delta H/2+\overline{Q}_m$ & 0 & $0$ & $-\overline{Q}_m$\\
    $\tilde{\Psi}^{\alpha}$ & $\delta H/2$ & $\delta H/2+\overline{Q}_m-1/2$ & 0 & $-1/2$ & $-\overline{Q}_m+1$ \\
    $\tilde{Y}$ & $\delta H/2$ & $\delta H/2+\overline{Q}_m-1$ & 0 & $-1$ & $-\overline{Q}_m+2$ \\ \hline
    \end{tabular}
    \caption{Charges of the massive particles and their mirror theory bound states under the Cartan elements.}
    \label{tab:mirrorcharges}
\end{table}

\subsection{Twisting the transfer matrix}
The twisted transfer matrix is, for each sector,
\begin{equation} 
    \TranB_\star^G(x^\pm) = \text{STr}_a \left[ G_{\star} S_{aN}(x^\pm,x_N^\pm) \dots S_{a1}(x^\pm,x_1^\pm)  \right]\,.
\end{equation}
In the above $S_{ab}(x^\pm,y^\pm)$ is the full, untwisted, S matrix. For instance, in the left sector,
 \begin{equation} \begin{aligned}
  \TranB_\L^G(x^\pm)  &= e^{i \mu_\L} \Big(e^{i \nu} \MA^{(1)}(x^\pm) \MA^{(2)}(x^\pm) - e^{\frac{i}{2} \dot{\rho}} \MA^{(1)}(x^\pm) \MD^{(2)}(x^\pm)\\
  &\qquad - e^{-\frac{i}{2} \dot{\rho}}  \MD^{(1)}(x^\pm) \MA^{(2)}(x^\pm) + e^{-i \nu} \MD^{(1)}(x^\pm) \MD^{(2)}(x^\pm)\Big)\,.
  \end{aligned}
\end{equation}
% The twist factorises, but with two different factors. This leads to two different Bethe equations for the two auxiliary roots
% \begin{equation}
%     \frac{\Omega_{\L,1}(y^{\pm,{(1)}})}{\Omega_{\L,2}(y^{\pm,{(1)}})} = e^{-i \nu} e^{-\frac{i}{2} \dot{\rho}}\,, \qquad  \frac{\Omega_{\L,1}(y^{\pm,{(2)}})}{\Omega_{\L,2}(y^{\pm,{(2)}})} = e^{-i \nu} e^{+\frac{i}{2} \dot{\rho}}\,.
% \end{equation}
Since the twist $G$ differs in each sector only through an overall factor of the type~$e^{i(\mu_\L-\mu_\R)}$, it will still be true that the various irreducible representations (left, right and massless) will only result in a single type of auxiliary Bethe root, just as in the untwisted case. However, while $G$ factorises, the action on each tensor product space --- corresponding to the split $\psu(1|1)^{\oplus4}\cong(\psu(1|1)^{\oplus2})^{\oplus2}$ --- is different, which leads to two different equations for the two sets of roots $\{y^{(1)}_j\}$ and $\{y^{(2)}_j\}$. These can be obtained from \eqref{eq:AuxBetheDelta2} by replacing
\begin{equation}
     \Delta_\star(y^{(\alpha)}) = \Omega_{\star,1}(y^{(\alpha)}) - \Omega_{\star,1}(y^{(\alpha)})\,, 
\end{equation}
     with two different functions 
\begin{align}
    \Delta^{(1)}_\star(y^{(1)})  &= e^{+\frac{i}{2} \nu+\frac{i}{2} \rho} \Omega_{\star,1}(y^{(1)}) - e^{-\frac{i}{2} \nu} \Omega_{\star,2}(y^{(1)})\,, \\
     \Delta^{(2)}_\star(y^{(2)})  &= e^{+\frac{i}{2} \nu-\frac{i}{2} \rho} \Omega_{\star,1}(y^{(2)}) - e^{-\frac{i}{2} \nu} \Omega_{\star,2}(y^{(2)})\,.
     \end{align}
The auxiliary Bethe equations are then
\begin{align}
e^{+i \nu} e^{+\frac{i}{2} \rho} &= \prod_{j=1}^{N_1} S^{y1}(y_i^{(1)},u_{j}) \prod_{j=1}^{N_{\bar{1}}} \bar{S}^{y1}(y_i^{(1)},u_{j})
\prod_{\dot{\alpha}=1}^2\prod_{j=1}^{N_0^{(\dot{\alpha})}} S^{y0}(y_i^{(1)},u_j^{(\dot{\alpha})})\,, \\
e^{+i \nu} e^{-\frac{i}{2} \rho} &= \prod_{j=1}^{N_1} S^{y1}(y_i^{(2)},u_{j}) \prod_{j=1}^{N_{\bar{1}}} \bar{S}^{y1}(y_i^{(2)},u_{j})
\prod_{\dot{\alpha}=1}^2\prod_{j=1}^{N_0^{(\dot{\alpha})}} S^{y0}(y_i^{(2)},u_j^{(\dot{\alpha})}).
\end{align}
Similarly, the eigenvalues of the transfer matrices can be obtained from \eqref{eq:eigDelta2} through the replacement (one now needs to make a distinction between the two massless particles)
\begin{align}
    (\Delta_\L(x^\pm))^2 &\rightarrow e^{i \mu_\L} \Delta^{(1)}_\L(x^\pm) \Delta^{(2)}_\L(x^\pm)\,, \\
    (\Delta_\R(x^\pm))^2 &\rightarrow e^{i \mu_\R} \Delta^{(1)}_\R(x^\pm) \Delta^{(2)}_\R(x^\pm)\,, \\
    (\Delta_\varoslash(x^\pm))^2 &\rightarrow e^{i \mu_\O} e^{+\frac{i}{2} \dot{\rho}} \Delta^{(1)}_\O(x^\pm) \Delta^{(2)}_\O(x^\pm)\,, \\
    (\Delta_\varobslash(x^\pm))^2 &\rightarrow e^{i \mu_\O} e^{-\frac{i}{2} \dot{\rho}} \Delta^{(1)}_\O(x^\pm) \Delta^{(2)}_\O(x^\pm)\,.
\end{align}
This leads to the eigenvalues
\begin{equation} \begin{aligned}
\mathbb{\Lambda}_\L(x^\pm)  &= e^{i \mu_\L} \prod_{j=1}^{N_1} 
S^{11}_{su}(x^\pm,u_j) \prod_{j=1}^{N_{\bar{1}}} \widetilde{S}^{11}_{su}(x^\pm,u_j) \prod_{\dot{\alpha}=1}^2 \prod_{j=1}^{N_0^{(\dot{\alpha})}} S^{10}(x^\pm,u_j^{(\dot{\alpha})}) \prod_{\alpha=1}^2 \prod_{j=1}^{N_y^{(\alpha)}} \overline{S}^{1y}(x^\pm,y^{(\alpha)}_j)\\
&\qquad 
\times \left(e^{\frac{i}{2} \nu+\frac{i}{2} \rho}-e^{-\frac{i}{2} \nu} \prod_{j=1}^{N_1}\overline{S}^{y1}(x^-,u_j) \prod_{j=1}^{N_{\bar{1}}}S^{y1}(x^-,u_j) \prod_{\dot{\alpha}=1}^2 \prod_{j=1}^{N_0^{(\dot{\alpha})}} S^{y0}(x^-,u_j^{(\dot{\alpha})}) \right) \\
&\qquad 
\times \left(e^{\frac{i}{2} \nu-\frac{i}{2} \rho}-e^{-\frac{i}{2} \nu} \prod_{j=1}^{N_1}\overline{S}^{y1}(x^-,u_j) \prod_{j=1}^{N_{\bar{1}}}S^{y1}(x^-,u_j) \prod_{\dot{\alpha}=1}^2 \prod_{j=1}^{N_0^{(\dot{\alpha})}} S^{y0}(x^-,u_j^{(\dot{\alpha})}) \right)\,,
\end{aligned}
\end{equation}
\begin{equation} \begin{aligned}
\mathbb{\Lambda}_\R(x^\pm)  &= e^{i \mu_\R} \prod_{j=1}^{N_1} 
\widetilde{S}^{11}_{sl}(x^\pm,u_j) \prod_{j=1}^{N_{\bar{1}}} S^{11}_{sl}(x^\pm,u_j) \prod_{\dot{\alpha}=1}^2 \prod_{j=1}^{N_0^{(\dot{\alpha})}} \bar{S}^{10}(x^\pm,u_j^{(\dot{\alpha})}) \prod_{\alpha=1}^2 \prod_{j=1}^{N_y^{(\alpha)}} S^{1y}(x^\pm,y^{(\alpha)}_j)  \\
&\qquad 
\times \left(e^{\frac{i}{2} \nu+\frac{i}{2} \rho}-e^{-\frac{i}{2} \nu} \prod_{j=1}^{N_1}\overline{S}^{y1}(\tfrac{1}{x^+},u_j) \prod_{j=1}^{N_{\bar{1}}}S^{y1}(\tfrac{1}{x^+},u_j) \prod_{\dot{\alpha}=1}^2 \prod_{j=1}^{N_0^{(\dot{\alpha})}} S^{y0}(\tfrac{1}{x^+},u_j^{(\dot{\alpha})}) \right) \\
&\qquad 
\times \left(e^{\frac{i}{2} \nu-\frac{i}{2} \rho}-e^{-\frac{i}{2} \nu} \prod_{j=1}^{N_1}\overline{S}^{y1}(\tfrac{1}{x^+},u_j) \prod_{j=1}^{N_{\bar{1}}}S^{y1}(\tfrac{1}{x^+},u_j) \prod_{\dot{\alpha}=1}^2 \prod_{j=1}^{N_0^{(\dot{\alpha})}} S^{y0}(\tfrac{1}{x^+},u_j^{(\dot{\alpha})}) \right)\,,
\end{aligned}
\end{equation}
\begin{equation} \begin{aligned}
\mathbb{\Lambda}_\varoslash(x^\pm)  &= - e^{i \mu_\O} e^{\frac{i}{2} \dot{\rho}} \prod_{j=1}^{N_1} 
S^{01}(x^\pm,u_j)  \prod_{j=1}^{N_{\bar{1}}} \bar{S}^{01}(x^\pm,u_j) \prod_{\dot{\alpha}=1}^2 \prod_{j=1}^{N_0^{(\dot{\alpha})}} S^{00}(x^\pm,u_j^{(\dot{\alpha})}) \prod_{\alpha=1}^2 \prod_{j=1}^{N_y^{(\alpha)}} S^{0y}(x^\pm,y^{(\alpha)}_j) \\
&\qquad 
\times \left(e^{\frac{i}{2} \nu+\frac{i}{2} \rho}-e^{-\frac{i}{2} \nu} \prod_{j=1}^{N_1}\overline{S}^{y1}(x^-,u_j) \prod_{j=1}^{N_{\bar{1}}}S^{y1}(x^-,u_j) \prod_{\dot{\alpha}=1}^2 \prod_{j=1}^{N_0^{(\dot{\alpha})}} S^{y0}(x^-,u_j^{(\dot{\alpha})}) \right) \\
&\qquad 
\times \left(e^{\frac{i}{2} \nu-\frac{i}{2} \rho}-e^{-\frac{i}{2} \nu} \prod_{j=1}^{N_1}\overline{S}^{y1}(x^-,u_j) \prod_{j=1}^{N_{\bar{1}}}S^{y1}(x^-,u_j) \prod_{\dot{\alpha}=1}^2 \prod_{j=1}^{N_0^{(\dot{\alpha})}} S^{y0}(x^-,u_j^{(\dot{\alpha})}) \right)\,,
\end{aligned}
\end{equation}
\begin{equation} \begin{aligned}
\mathbb{\Lambda}_\varobslash(x^\pm)  &= - e^{i \mu_\O} e^{-\frac{i}{2} \dot{\rho}} \prod_{j=1}^{N_1} 
S^{01}(x^\pm,u_j)  \prod_{j=1}^{N_{\bar{1}}} \bar{S}^{01}(x^\pm,u_j) \prod_{\dot{\alpha}=1}^2 \prod_{j=1}^{N_0^{(\dot{\alpha})}} S^{00}(x^\pm,u_j^{(\dot{\alpha})}) \prod_{\alpha=1}^2 \prod_{j=1}^{N_y^{(\alpha)}} S^{0y}(x^\pm,y^{(\alpha)}_j) \\
&\qquad 
\times \left(e^{\frac{i}{2} \nu+\frac{i}{2} \rho}-e^{-\frac{i}{2} \nu} \prod_{j=1}^{N_1}\overline{S}^{y1}(x^-,u_j) \prod_{j=1}^{N_{\bar{1}}}S^{y1}(x^-,u_j) \prod_{\dot{\alpha}=1}^2 \prod_{j=1}^{N_0^{(\dot{\alpha})}} S^{y0}(x^-,u_j^{(\dot{\alpha})}) \right) \\
&\qquad 
\times \left(e^{\frac{i}{2} \nu-\frac{i}{2} \rho}-e^{-\frac{i}{2} \nu} \prod_{j=1}^{N_1}\overline{S}^{y1}(x^-,u_j) \prod_{j=1}^{N_{\bar{1}}}S^{y1}(x^-,u_j) \prod_{\dot{\alpha}=1}^2 \prod_{j=1}^{N_0^{(\dot{\alpha})}} S^{y0}(x^-,u_j^{(\dot{\alpha})}) \right)\,.
\end{aligned}
\end{equation}

The main Bethe equations are
\begin{equation} \begin{aligned}
1 &= e^{i p_i L} e^{i \mu_\L} \prod_{j\neq i}^{N_1} 
S^{11}_{su}(u_i,u_j) \prod_{j=1}^{N_{\bar{1}}} \widetilde{S}^{11}_{su}(u_i,u_j) \prod_{\dot{\alpha}=1}^2 \prod_{j=1}^{N_0^{(\dot{\alpha})}} S^{10}(u_i,u_j^{(\dot{\alpha})}) \prod_{\alpha=1}^2 \prod_{j=1}^{N_y^{(\alpha)}} \overline{S}^{1y}(u_i,y^{(\alpha)}_j)\,,
\end{aligned}
\end{equation}
\begin{equation} \begin{aligned}
1 &= e^{i p_i L}  e^{i \mu_\R} \prod_{j=1}^{N_1} 
\widetilde{S}^{11}_{sl}(u_i,u_j) \prod_{j \neq i}^{N_{\bar{1}}} S^{11}_{sl}(u_i,u_j) \prod_{\dot{\alpha}=1}^2 \prod_{j=1}^{N_0^{(\dot{\alpha})}} \bar{S}^{10}(u_i,u_j^{(\dot{\alpha})}) \prod_{\alpha=1}^2 \prod_{j=1}^{N_y^{(\alpha)}} S^{1y}(u_i,y^{(\alpha)}_j)\,,
\end{aligned}
\end{equation}
\begin{equation} \begin{aligned}
1 &= e^{i p_i L} e^{i \mu_\O} e^{\frac{i}{2} \dot{\rho}} \prod_{j=1}^{N_1} 
S^{01}(x^\pm,u_j)  \prod_{j=1}^{N_{\bar{1}}} \bar{S}^{01}(x^\pm,u_j) \prod_{\dot{\alpha}=1}^2 \prod_{j=1}^{N_0^{(\dot{\alpha})}} S^{00}(x^\pm,u_j^{(\dot{\alpha})}) \prod_{\alpha=1}^2 \prod_{j=1}^{N_y^{(\alpha)}} S^{0y}(x^\pm,y^{(\alpha)}_j)\,.
\end{aligned}
\end{equation}
\begin{equation} \begin{aligned}
1 &= e^{i p_i L} e^{i \mu_\O} e^{-\frac{i}{2} \dot{\rho}} \prod_{j=1}^{N_1} 
S^{01}(x^\pm,u_j)  \prod_{j=1}^{N_{\bar{1}}} \bar{S}^{01}(x^\pm,u_j) \prod_{\dot{\alpha}=1}^2 \prod_{j=1}^{N_0^{(\dot{\alpha})}} S^{00}(x^\pm,u_j^{(\dot{\alpha})}) \prod_{\alpha=1}^2 \prod_{j=1}^{N_y^{(\alpha)}} S^{0y}(x^\pm,y^{(\alpha)}_j)\,.
\end{aligned}
\end{equation}

\section{Conclusions}
\label{sec:conclusions}
We used the algebraic Bethe ansatz to (re)derive the Bethe equations for the $AdS_3\times S^3\times T^4$ string and mirror models, and to derive the eigenvalues of the string and mirror bound-state transfer matrices. We took some care to analyse a particular feature of this model and we showed that, even if the theory features several irreducible representations for fundamental particles, this does not result in the existence of several distinct auxiliary roots. Finally, we have extended our discussions to the case of Abelian twists that commute with the light-cone Hamiltonian.

The results of this paper will be useful for a detailed exploration of the spectrum of the model using the mirror thermodynamic Bethe ansatz of~\cite{Frolov:2021bwp}. Moreover, it would be interesting to study the twisted model in relation to TsT deformations of the string models. These have been considered for the related WZW models~\cite{Forste:2003km}, also in relation to $T\bar{T}$ deformations~\cite{Apolo:2019zai,Apolo:2021wcn}, see also~\cite{Giveon:2017nie,Giveon:2017myj}. It would be interesting to understand what happens for RR backgrounds. On a related note, twisted boundary conditions can be used to define ``fishnet'' models, like it was done for $AdS_5\times S^5$~\cite{Gurdogan:2015csr} and for~$AdS_4\times CP^3$~\cite{Caetano:2016ydc}. There, a large part of the appeal of the resulting models was the existence of a simple local Lagrangian for the dual CFT, string from $\mathcal{N}=4$ SYM and ABJM, respectively. Clearly, things will be more complicated in the AdS3/CFT2 setting, as the dual of the pure-RR theory is expected to be non-local, see~\cite{OhlssonSax:2014jtq} for a discussion of its weak-coupling structure. Still, perhaps a strongly-twisted ``fishnet'' limit exists, and might simplify some of the features of the model.
We hope to return to some of these questions in the future.

\section*{Acknowledgements}
We thank Marius de Leeuw and Sergey Frolov for useful related discussions and comments on the manuscript. 
The work of FS is supported by the Swiss National Science
Foundation via the Early Postdoc.Mobility fellowship ``q-deforming AdS/CFT''.
AS gratefully acknowledges support from the IBM Einstein Fellowship.

\appendix
\section{Bound states and Fermion grading}
\label{app:boundstates}

Let us briefly review the construction of bound states in the symmetric and anti-symmetric representation (see also~\cite{Majumder:2021zkr}).
Here we will consider the two-particle representation~\eqref{eq:twoparticlerepr} in the case where the two momenta satisfy the bound-state condition. We will then see that the representation becomes reducible (but indecomposable). A short two-dimensional representation for the bound states can then be constructed by quotienting out the indecomposable representation by a subrepresentation~\cite{Arutyunov:2008zt}. This will allow us to check which representation is symmetric and which is antisymmetric (in the string and mirror theory). We summarise here our findings:
\begin{itemize}
    \item The string-theory bound states are in the symmetric representation. For LL bound representations, the highest-weight state is bosonic, whereas for RR representations it is fermionic.
    \item The mirror-theory bound states are in the anti-symmetric representation. For LL bound representations, the highest-weight state is fermionic, whereas for RR representations it is bosonic.
\end{itemize}

\subsection{LL bound states}
We work in the basis $(\phi_\L^\B \otimes \phi_\L^\B,\phi_\L^\B \otimes \varphi_\L^\F,\varphi_\L^\F \otimes \phi_\L^\B,\varphi_\L^\F \otimes \varphi_\L^\F)$. Let $\gen{J}$ be any of the supercharges in the left-left two-particle representation defined by~\eqref{eq:twoparticlerepr}. It is convenient to make the two-particle change of basis~\cite{Majumder:2021zkr}
\begin{equation}
\gen{J}' = \gen{F}^{-1} \gen{J} \gen{F}~,
\end{equation}
with
\begin{equation}
\gen{F} = \frac{1}{n}\begin{pmatrix}
n & 0 & 0 & 0 \\
0 & e^{\frac{i p_1}{2}}\eta_2 & - e^{\frac{-i p_1}{2}}\eta_1& 0 \\
0 &  \eta_1& e^{-\frac{i (p_1+p_2)}{2}}\eta_2 & 0 \\
0 & 0 & 0 & n
\end{pmatrix}, \quad n=\sqrt{e^{\frac{-i p_1}{2}}\eta_1^2 + e^{-\frac{i p_2}{2}}\eta_2^2}.
\end{equation}
Then in the new basis we have
\begin{equation}
\gen{Q}' = n \begin{pmatrix}
0 & 0 & 0 & 0 \\
1 & 0 & 0 & 0 \\
0 & 0 & 0 & 0 \\
0 & 0 & -1 & 0 
\end{pmatrix}, \qquad \gen{S}' = n \begin{pmatrix}
0 & 1 & 0 & 0 \\
0 & 0 & 0 & 0 \\
0 & 0 & 0 & -1 \\
0 & 0 & 0 & 0 
\end{pmatrix},
\end{equation}
as well as
\begin{equation}
\widetilde{\gen{S}}' = - \frac{\eta_1 \eta_2}{x^+_1 x^+_2}\frac{1}{n}
\begin{pmatrix}
0 & 0 & 0 & 0\\
s' & 0 & 0 & 0 \\
e^{\frac{i p_1}{2}}(x^-_1-x^+_2) & 0 & 0 & 0\\
0 & e^{\frac{i p_1}{2}}(x^-_1-x^+_2) & -s'  & 0
\end{pmatrix}, 
\end{equation}
and
\begin{equation}
\widetilde{\gen{Q}}' = -\frac{\eta_1 \eta_2}{x^-_1 x^-_2}\frac{1}{n}
\begin{pmatrix}
0 & q' & e^{\frac{-i (2p_1+p_2)}{2}}(x^+_1-x^-_2) & 0\\
0  & 0 & 0 & e^{\frac{i (2 p_1+p_2)}{2}}(x^+_1-x^-_2) \\
0 & 0 & 0 & -q'\\
0 & 0 & 0  & 0
\end{pmatrix}.
\end{equation}
In the above
\begin{align}
    s' =e^{\frac{-i p_1}{2}}x^+_2 \frac{\eta_1}{\eta_2} +e^{\frac{-i p_2}{2}}x^-_1 \frac{\eta_2}{\eta_1}~, \qquad 
    q' = e^{\frac{-i p_1}{2}}x^-_2 \frac{\eta_1}{\eta_2} +e^{\frac{-i p_2}{2}}x^+_1 \frac{\eta_2}{\eta_1}~.
\end{align}
The S-matrix in the new basis, $S'(p_1,p_2)=F^{-1}(p_2,p_1)S(p_1,p_2)F(p_1,p_2)$, becomes diagonal
\begin{equation}
    S'_{12} = \Sigma^\LL_{12} \begin{pmatrix}
    1 & 0 & 0 & 0 \\
    0 & 1 & 0 & 0 \\
    0 & 0 & -e^{\frac{-i(p_1-p_2)}{2}} \frac{x^+_1-x^-_2}{x^-_1-x^+_2} & 0 \\
    0 & 0 & 0 & -e^{\frac{-i(p_1-p_2)}{2}} \frac{x^+_1-x^-_2}{x^-_1-x^+_2} 
    \end{pmatrix}~.
\end{equation}

\paragraph{String-theory bound state.}
In the string theory the bound state condition is instead~\cite{Frolov:2021fmj}
\begin{equation}
x^+(p_1) = x^-(p_2)\,,
\end{equation}
and we set
\begin{equation}
     X^-(p)=x^-(p_1)\,, \qquad X^+(p)=x^+(p_2)\,.
\end{equation}
Then
\begin{equation}
\widetilde{\gen{S}}' =  - \frac{n}{ X^+(p)} \begin{pmatrix}
0 & 0 & 0 & 0 \\
1 & 0 & 0 & 0 \\
\#& 0 & 0 & 0 \\
0 & \# & -1 & 0 
\end{pmatrix}~,
\qquad
\widetilde{\gen{Q}}' = -\frac{n}{X^-(p)} \begin{pmatrix}
0 & 1 & 0 & 0 \\
0 & 0 & 0 & 0 \\
0 & 0 & 0 & -1 \\
0 & 0 & 0 & 0 
\end{pmatrix}~.
\end{equation}
The $2 \times 2$ lower right blocks form a two-dimensional subrepresentation. The bound-state representation is found by quotienting the original four-dimensional indecomposable representation by this two-dimensional subrepresentation~\cite{Arutyunov:2008zt}. This block is spanned by the state $\phi_\L^\B \otimes \phi_\L^\B$ and by a symmetric combination of $\phi_\L^\B \otimes \varphi_\L^\F$ and $\varphi_\L^\F\otimes \phi_\L^\B $. Hence, we are dealing with a symmetric representation with \textit{bosonic highest-weight state}.
(Note that this representation was mis-identified in~\cite{Majumder:2021zkr}.)

\paragraph{Mirror-theory bound state.}
In the mirror theory, we see that we may create a bound-state in the Bethe equations when
\begin{equation}
x^-(p_1) = x^+(p_2).
\end{equation}
Let us also define the bound-state parameters
\begin{equation}
    X^+(p)=x^+(p_1)\,, \qquad  X^-(p)=x^-(p_2)\,.
\end{equation}
We then have $n = e^{\frac{-ip}{4}}\eta(p)$ and
\begin{equation}
\widetilde{\gen{S}}' = -\frac{n}{X^+(p)} \begin{pmatrix}
0 & 0 & 0 & 0 \\
1 & 0 & 0 & 0 \\
0 & 0 & 0 & 0 \\
0 & 0 & -1 & 0 
\end{pmatrix}~, \qquad
\widetilde{\gen{Q}}' = -\frac{n}{X^-(p)}\begin{pmatrix}
0 & 1 & \# & 0 \\
0 & 0 & 0 & \# \\
0 & 0 & 0 & -1\\
0 & 0 & 0 & 0 
\end{pmatrix}, 
\end{equation}
where ``$\#$'' indicates some non-zero terms.
From this it is easy to see that the $2 \times 2$ upper left blocks form a sub-representation. Once again, the bound-state representation is found by factoring the original four-dimensional indecomposable representation by this sub-representation.
Hence, we are dealing with an anti-symmetric bound-state representation with \textit{fermionic highest-weight state} given by a linear combination of $\phi_\L^\B \otimes \varphi_\L^\F$ and $\varphi_\L^\F\otimes \phi_\L^\B $.

\subsection{RR bound states}
We work in the basis $(\phi_\R^\F \otimes \phi_\R^\F,\phi_\R^\F \otimes \varphi_\R^\B,\varphi_\R^\B \otimes \phi_\R^\F,\varphi_\R^\B \otimes \varphi_\R^\B)$ and make the two-particle change of basis
\begin{equation}
\gen{J}' = \gen{F}^{-1} \gen{J} \gen{F}\,,
\end{equation}
with
\begin{equation}
\gen{F} = \frac{1}{n}\begin{pmatrix}
n & 0 & 0 & 0 \\
0 & -e^{-\frac{i (p_1+p_2)}{2}}\eta_2 & -\eta_1& 0 \\
0 &  e^{\frac{-i p_1}{2}}\eta_1 & -e^{\frac{i p_1}{2}}\eta_2 & 0 \\
0 & 0 & 0 & n
\end{pmatrix}, \qquad 
n=\sqrt{e^{\frac{-i p_1}{2}}\eta_1^2 + e^{-\frac{i p_2}{2}}\eta_2^2}\,.
\end{equation}
Then in the new basis we have
\begin{equation}
\tilde{\gen{S}}' = n \begin{pmatrix}
0 & 0 & 0 & 0 \\
1 & 0 & 0 & 0 \\
0 & 0 & 0 & 0 \\
0 & 0 & -1 & 0 
\end{pmatrix}~, \qquad \tilde{\gen{Q}}' = n \begin{pmatrix}
0 & 1 & 0 & 0 \\
0 & 0 & 0 & 0 \\
0 & 0 & 0 & -1 \\
0 & 0 & 0 & 0 
\end{pmatrix}~.
\end{equation}
The generators $\gen{Q}'$ and $\gen{S}'$ are more involved, we only need their explicit expressions in the two cases discussed below.

\paragraph{String-theory bound state.} If we now set 
\begin{equation}
x^+(p_1) = x^-(p_2)\,,
\end{equation}
and define
\begin{equation}
     X^-(p)=x^-(p_1)\,, \qquad X^+(p)=x^+(p_2)\,,
\end{equation}
then we have
\begin{equation}
\gen{Q}' =  - \frac{n}{X^-(p)} \begin{pmatrix}
0 & 0 & 0 & 0 \\
1 & 0 & 0 & 0 \\
0& 0 & 0 & 0 \\
0 & 0 & -1 & 0 
\end{pmatrix},
\qquad
\gen{S}' = -\frac{n}{X^+(p)} \begin{pmatrix}
0 & 1 & \# & 0 \\
0 & 0 & 0 & \# \\
0 & 0 & 0 & -1 \\
0 & 0 & 0 & 0 
\end{pmatrix}.
\end{equation}
The $2 \times 2$ upper left blocks form a subrepresentation. The bound-state representation is obtained by quotient, as described above.
This representation spans the vector space consisting of the symmetric combination of $\phi_\R^\F \otimes \varphi_\R^\B$ and $\varphi_\R^\B \otimes \phi_\R^\F$, as well as $\varphi_\R^\B \otimes \varphi_\R^\B$. Hence, we are dealing with a symmetric representation with \textit{fermionic highest-weight state}.

\paragraph{Mirror-theory bound state.} If we set 
\begin{equation}
x^-(p_1) = x^+(p_2)\,,
\end{equation}
and define
\begin{equation}
X^+(p)=x^+(p_1)\,, \qquad X^-(p)=x^-(p_2),
\end{equation}
then $n = e^{\frac{-ip}{4}}\eta(p)$ and
\begin{equation}
\gen{Q}' = -\frac{n}{X^-(p)} \begin{pmatrix}
0 & 0 & 0 & 0 \\
1 & 0 & 0 & 0 \\
\# & 0 & 0 & 0 \\
0 & \# & -1 & 0 
\end{pmatrix}, \qquad
\gen{S}' = -\frac{n}{X^+(p)}\begin{pmatrix}
0 & 1 & 0 & 0 \\
0 & 0 & 0 & 0 \\
0 & 0 & 0 & -1\\
0 & 0 & 0 & 0 
\end{pmatrix}\,.
\end{equation}
The $2 \times 2$ lower right blocks form a sub-representation and the bound-state representation is found by quotient.
Hence, we are dealing with a anti-symmetric representation with \textit{bosonic highest-weight state}.

\section{All creation operators are created equal}
\label{app:Null}
In this appendix we demonstrate that all states 
\begin{equation}
    \MB_\L(y_\L^\pm) \ket{0}\, \qquad \MB_\R(y^\pm_\R) \ket{0}\, \qquad \MB_\o(y^\pm_\o) \ket{0}\,,\qquad \MB_\op(y^\pm_\op) \ket{0}\,,
\end{equation}
are proportional to each other provided that 
\begin{equation}
\label{eq:yrel}
 y_\L^\pm=y^\pm_\o=y^\pm_\op=1/y^\mp_\R\,.
\end{equation}
To achieve this we show that there exist linear combinations
\begin{equation}
 \ket{\psi}=\MB_\star(y^\pm_\star) \ket{0}+ c(y^\pm_\star,y^\pm_{\star'}) \MB_{\star'}(y^\pm_{\star'}) \ket{0}\,,
\end{equation}
whose norm is zero for appropriately chosen coefficients $c(y^\pm_\star,y^\pm_{\star'})$. Since
\begin{equation}
\bra{\psi} =\bra{0}\ \MC_\star\left((y^\pm_\star)^*\right) + \left(c(y^\pm_\star,y^\pm_{\star'})\right)^* \bra{0}\  \MC_{\star'}\left((y^\pm_{\star'})^*\right) \,,
\end{equation}
this entails computing quantities of the form $\bra{0} \MC_\star(y^\pm_{1}) \MB_{\star'}(y_{2}^\pm) \ket{0}$. In order to avoid singularities we assume two different rapidities $y^\pm_{1}$ and $y^\pm_{2}$ and only take the appropriate limit at the end of the calculation. 

Let us focus on the linear combination involving left and right creation operators, which is the most nontrivial (recall that massless modes can be obtained as limits of the massive ones). We consider the linear combination
\begin{equation}
    \ket{\psi} = \MB_\L(y^\pm) \ket{0}+ c(y^\pm)\, \MB_\R(1/y^{\mp}) \ket{0}
\end{equation}
and aim to find the function $c(y^\pm)$ such that this state has zero norm.

From the RTT relations it follows that (to simplify the result we use the explicit expression of $\eta(y^\pm)$)
\begin{align}
    \bra{0} \MC_\L(y_1^\pm) \MB_\L(y_2^\pm) \ket{0} &=  \sqrt{\frac{y_1^-}{y_1^+}} \sqrt{\frac{y_2^-}{y_2^+}}  \,\Omega_{\L,1}(y_1^\pm) \Omega_{\L,2}(y_2^\pm) \, \frac{2i}{h}\eta_1 \eta_2\frac{\Upsilon (y_1^-,y_2^-)}{{y_1^--y^-_2}}\,, \\
    \bra{0} \MC_\L(y_1^\pm) \MB_\R(1/y_2^\mp) \ket{0} &= \sqrt{\frac{y_1^-}{y_1^+}}  \frac{y_2^-}{y_2^+}  \, \Omega_{\L,1}(y_1^\pm) \Omega_{\R,2}(1/y_2^\mp) \, \frac{2 i}{h} \eta_1 \eta_2\frac{\Upsilon (y_1^-,y_2^-)}{{y^-_1-y_2^-}}\,, \\
    \bra{0} \MC_\R(1/y_1^\mp) \MB_\L(y_2^\pm) \ket{0} &= \frac{y_1^-}{y_1^+} \sqrt{\frac{y_2^-}{y_2^+}}  \,\Omega_{\R,1}(1/y_1^\mp) \Omega_{\L,2}(y_2^\pm) \,  \frac{2i}{h} \eta_1 \eta_2\frac{\Upsilon (y_1^-,y_2^-)}{y_1^--y_2^-}\,, \\
    \bra{0} \MC_\R(1/y_1^\mp)\MB_\R(1/y_2^\mp) \ket{0} &= \frac{y_1^-}{y_1^+} \frac{y_2^-}{y_2^+}  \, \Omega_{\R,1}(1/y_1^\mp) \Omega_{\R,2}(1/y_2^\mp) \, \frac{2 i}{h}  \eta_1 \eta_2\frac{\Upsilon (y_1^-,y_2^-)}{y_1^--y_2^-}\,,
\end{align}
% \begin{align}
%     \bra{0} \MC_\L(y_1^\pm) \MB_\L(y_2^\pm) \ket{0} &= - e^{-\frac{i q_1 }{2}} \frac{y^-_2 - y^+_2}{y_1^--y^-_2} \frac{\eta_1}{\eta_2} \Omega_{\L,1}(y_1^\pm) \Omega_{\L,2}(y_2^\pm) \Upsilon (y_1^-,y_2^-)\,, \\
%     \bra{0} \MC_\L(y_1^\pm) \MB_\R(y_2^\pm) \ket{0} &= e^{-\frac{iq_1}{2}} \frac{2 i}{h} \frac{1}{y^+_2}\frac{\eta_1 \eta_2}{y^-_1-\frac{1}{y^+_2}} \Omega_{\L,1}(y_1^\pm) \Omega_{\R,2}(y_2^\pm) \Upsilon (y_1^-,1/y_2^+)\,, \\
%     \bra{0} \MC_\R(y_1^\pm) \MB_\L(y_2^\pm) \ket{0} &= e^{-\frac{i q_2}{2}}\frac{2i}{h} \frac{1}{y_1^+} \frac{\eta_1 \eta_2}{\frac{1}{y_1^+}-y_2^-} \Omega_{\R,1}(y_1^\pm) \Omega_{\L,2}(y_2^\pm) \Upsilon (1/y_1^+,y_2^-)\,, \\
%     \bra{0} \MC_\R(y_1^\pm) \MB_\R(y_2^\pm) \ket{0} &= -e^{-\frac{i q_2}{2}} \frac{1}{y_1^+} \frac{1-\frac{y_2^+}{y_2^-}}{\frac{1}{y_1^+}-\frac{1}{y^+_2}} \frac{\eta_1}{\eta_2} \Omega_{\R,1}(y_1^\pm) \Omega_{\R,2}(y_2^\pm) \Upsilon (1/y_1^+,1/y_2^+)\,,
% \end{align}
with 
\begin{equation}
    \Upsilon (y_1,y_2)= 1-\prod_{\mathrm{J}}\prod_{j=1}^{L_\mathrm{J}} S^\mathrm{II,I}_{\cdot \mathrm{J}} (y_1,x^\pm_j) S^\mathrm{I,II}_{\mathrm{J} \cdot} (x^\pm_j,y_2)\,.
\end{equation}
One is then interested in the limit $y_{1}^\pm \rightarrow (y^\pm)^*$ and $y_{2}^\pm \rightarrow y^\pm$. Notice that
\begin{equation}
    \Upsilon (y_1,y_2)=(y_1-y_2) \hat{\Upsilon}(y_1,y_2)~,
\end{equation}
where $\hat{\Upsilon}(y_1,y_2)$ has no singularity in that limit. It then follows that upon taking the limit  the norm of $\ket{\psi}$ vanishes provided that
\begin{equation}
    c(y^\pm) = - \sqrt{\frac{y^+}{y^-}}\frac{\Omega_{\L,2}(y^\pm)}{\Omega_{\R,2}(1/y^\mp)}\,.
\end{equation}

Similar results are obtained for the two remaining linear combinations $\MB_\o(y^\pm) \ket{0}+ c(y^\pm) \MB_\R(1/y^\mp) \ket{0}$ and $\MB_\op(y^\pm) \ket{0}+ c(y^\pm) \MB_\R(1/y^\mp) \ket{0}$. 
%This shows that exited states constructed with different creation operators $\MB_\star(y)$ are all proportional to each other, so it is conceptually coherent to assume that there is only one type of creation operator $\MB(y)$.

Finally, let us mention that this proportionality argument naturally extends to states for which the operators $\mathcal B_\L(y^\pm)$ and $\mathcal B_\R(1/y^\mp)$ are part of strings of $\mathcal B$ operators, as in \eqref{eq:severalexfact}, rather than just acting on the vacuum.
For instance, let us consider two states of the form
\begin{equation}
    \MB_\L(y_\L^\pm)\prod_j\MB_{\star}(y^\pm_{\star,j})  \ket{0}\,, \qquad \MB_\R(y^\pm_\R) \prod_j\MB_{\star}(y^\pm_{\star,j})\ket{0}\,,
\end{equation}
where the products are over two identical (arbitrary) strings of $\MB$-operators, and the rapidities of the standalone operators satisfy $y_\L^\pm=1/y_\R^\mp$.
One can use the commutation relations \eqref{eq:BBi}--\eqref{eq:BBf} to bring the standalone $\MB$-operators to the right, so that they act on the vacuum.
This yields some scalar prefactor. For example, we have
\begin{align}
    \mathcal B_\L (y^\pm) \mathcal B_\L (x^\pm) &= \frac{F^\LL(y^\pm,x^\pm)}{A^\LL(y^\pm,x^\pm)} \mathcal B_\L(x^\pm) \mathcal B_\R (y^\pm)~, \\
    \mathcal B_\R (1/y^\mp) \mathcal B_\L (x^\pm) &= -\frac{C^\LR(x^\pm,1/y^\mp)}{D^\LR(x^\pm,1/y^\mp)} \mathcal B_\L(x^\pm) \mathcal B_\R (1/y^\mp)\,.
\end{align}
In fact, using the explicit expressions of the coefficients $F^\LL$, $A^\LL$, $C^\LR$ and $D^\LR$ we find that the prefactors are identical, \textit{e.g.}
\begin{equation}
    \frac{F^\LL(y^\pm,x^\pm)}{A^\LL(y^\pm,x^\pm)} =-\frac{C^\LR(x^\pm,1/y^\mp)}{D^\LR(x^\pm,1/y^\mp)}\,,
\end{equation}
and similarly for other S-matrix elements.
Therefore, calling the product over the reordering factors~$C(y_\L^\pm;y_{\star,j})$, the two expressions are now of the form
\begin{equation}
\begin{aligned}
    &C(y_\L^\pm;y_{\star,j})\left(\prod_j\MB_{\star,j}(y_{\star,j})\right)\MB_\L(y_\L^\pm)  \ket{0}=C(y_\L^\pm;\{y_{\star,j}\})\left(\prod_j\MB_{\star,j}(y_{\star,j})\right)|\Psi\rangle\,, \\ 
    &C(y_\L^\pm;y_{\star,j})\left( \prod_j\MB_{\star,j}(y_{\star,j})\right)\MB_\R(y^\pm_\R)\ket{0}=C(y_\L^\pm;\{y_{\star,j}\})\left( \prod_j\MB_{\star,j}(y_{\star,j})\right)c\ket{\Psi}\,,
\end{aligned}
\end{equation}
where in the last equality we used the fact, proved above, that the two standalone $\MB$ operators produce proportional states. This construction generalises to the case of multiple equivalent roots, as well as the case in which the string of $\MB$ operators (the product over~$j$) is replaced by a linear combination of products of $\MB$ operators.
The case where one acts with $\mathcal B_\o$ or $\mathcal B_\op$ follows the same logic.

\bibliographystyle{JHEP}
\bibliography{refs}

\end{document}